\documentclass{aa}

\usepackage{epsfig}
\usepackage{graphics}
\usepackage{natbib}

\usepackage{amssymb}
\usepackage{amsmath}

\newcommand{\h}{\bar{H}}
\newcommand{\x}{\bar{x}}
\newcommand{\y}{\bar{y}}

\newcommand{\Planck}{\textit{Planck }}

\usepackage{url}

\defcitealias{bib:neveu}{N13}   
\defcitealias{bib:neveu14}{N14}         
\defcitealias{bib:barreira13}{B13a}     
\defcitealias{bib:barreira14}{B14}      

\begin{document}

\title{Constraining the $\Lambda$CDM and Galileon models with recent cosmological data}
\author{J.~Neveu\inst{1,2,3}, V.~Ruhlmann-Kleider\inst{3}, P.~Astier\inst{4}, M.~Besan\c{c}on\inst{3}, J.~Guy\inst{4}, A.~M\"oller\inst{5,6,3}, E. Babichev\inst{7}
}
\institute{LAL, Univ. Paris-Sud, CNRS/IN2P3, Universit\'e Paris-Saclay, Orsay, France 
\and LERMA, UMR 8112 du CNRS, Observatoire de Paris, \'Ecole Normale Sup\'erieure, 24 rue Lhomond, F75231 Paris Cedex 05, France
\and CEA, Centre de Saclay, Irfu/SPP,  91191 Gif-sur-Yvette, France
\and  LPNHE, Universit\'e Pierre et Marie Curie, Universit\'e Paris Diderot, CNRS-IN2P3, 4 place Jussieu, 75252 Paris Cedex 05, France
\and Research School of Astronomy and Astrophysics, Australian National University, Canberra, ACT 2611, Australia
\and ARC Centre of Excellence for All-sky Astrophysics (CAASTRO), Australia
\and Laboratoire de Physique Th\'eorique, CNRS, Univ. Paris-Sud, Universit\'e Paris-Saclay, 91405 Orsay, France
}

\date{\today}
\authorrunning{J. Neveu et al.}
\titlerunning{Constraining the $\Lambda$CDM and Galileon models with recent cosmological data}

\abstract{}{The Galileon theory belongs to the class of modified gravity models that can explain the late-time accelerated expansion of the Universe. In previous works, cosmological constraints on the Galileon model were derived, both in the uncoupled case and with a disformal coupling of the Galileon field to matter. 
There, we showed that these models agree with the most recent cosmological data. In this work, we used updated cosmological data sets to derive new constraints on Galileon models, including the case of a constant conformal Galileon coupling to matter.
 We also explored the tracker solution of the uncoupled Galileon model.}{After updating our data sets, especially with the latest \textit{Planck} data and baryonic acoustic oscillation (BAO) measurements, we fitted the cosmological parameters of the $\Lambda$CDM and Galileon models. The same analysis framework as in our previous papers was used to derive cosmological constraints, using precise measurements of cosmological distances and of the cosmic structure growth rate.}{We show that all tested Galileon models are as compatible with cosmological data as the $\Lambda$CDM model. This means that present cosmological data are not accurate enough to distinguish clearly between the two theories. 
Among the different Galileon models, we find that a conformal coupling is not favoured, contrary to the disformal coupling which is preferred at the $2.3\sigma$ level over the uncoupled case. The tracker solution of the uncoupled Galileon model is also highly disfavoured owing to large tensions with supernovae and \textit{Planck}+BAO data. However, outside of the tracker solution, the general uncoupled Galileon model, as well as the general disformally coupled Galileon model, remain the most promising Galileon scenarios to confront with future cosmological data. Finally, we also discuss constraints coming from the Lunar Laser Ranging experiment and gravitational wave speed of propagation.}{}
\keywords{Cosmology: dark energy - Cosmology: observations - Cosmology: theory}
\maketitle

\section{Introduction}

Since the discovery of the accelerated expansion of the Universe \citep{bib:riess,bib:perlmutter}, the exact nature of dark energy is still unknown. The Einstein cosmological constant $\Lambda$ is the simplest way to describe the expansion measurements. The so-called $\Lambda$CDM model is the model that best agrees with actual cosmological data (see e.g. \citealt{bib:planck15}). However, the physical justification behind the cosmological constant is still under question. Despite a very good agreement with data, alternatives to the $\Lambda$CDM model have been introduced to escape the theoretical difficulties raised by the cosmological constant.

The Galileon theory \citep{bib:nicolis} belongs to the class of modified gravity models which aims to give an alternative explanation to the nature of dark energy. It introduces a scalar field, hereafter called $\pi$, whose equation of motion must be of second order and invariant under a Galilean shift symmetry $\partial_\mu \pi \rightarrow \partial_\mu \pi + b_\mu$, where $b_\mu$ is a constant vector. This symmetry was first identified as an interesting property in the DGP model \citep{bib:dgp}. \cite{bib:nicolis} derived the five possible Lagrangian terms for the $\pi$ field,  which were then formulated in a covariant formalism by \cite{bib:deffayet,bib:deffayetb}. The phenomenology of the Galileon theory was  then studied in  e.g.  \citet{bib:gannouji} and \citet{bib:appleby}.

This model forms a subclass of general tensor-scalar theories involving only up to second-order derivatives originally found by Horndeski \citep{bib:horndeski}. Later, subsets of Galileon theory were also found to be the decoupling limit of numerous broader theories, such as massive gravity \citep{bib:deRhamMasGra,bib:deRhamMasGra2} or brane constructions \citep{bib:deRhamDBI,bib:hinterbichler,bib:acoleyen}. Braneworld approaches give a deeper theoretical basis to Galileon theories. The usual and simple construction involves a 3+1 dimensional brane, our Universe, embedded in a higher dimensional bulk. The Galileon field $\pi$ can be interpreted as the brane transverse position in the bulk, and the Galilean symmetry appears naturally as a remnant of the broken space-time symmetries of the bulk \citep{bib:hinterbichler}. The Galilean symmetry is then no longer imposed as a principle of construction, but is a consequence of space-time geometry. 

Models that modify general relativity have to alter gravity only at cosmological scales in order to agree with the solar system tests of gravity (see e.g. \citealt{bib:will}). The Galileon field can be coupled to matter either explicitly or through a coupling induced by its temporal variation \citep{bib:babichev13}. This leads to a so-called fifth force, which by definition modifies gravity around massive objects like the Sun. But the non-linear nature of the Galileon theory ensures that this fifth force is screened near massive objects in case of an explicit conformal coupling of the form $\sim \pi T^\mu_{\ \mu}$ (where $T^\mu_{\ \mu}$ is the trace of the matter energy-momentum tensor) or in the case of an induced coupling. This is called the Vainshtein effect (\cite{bib:vainshtein} and \cite{bib:babichev13b} for a modern introduction). The fifth force is thus negligible with respect to general relativity within a certain radius from a massive object that depends on the object mass and on the values of the Galileon parameters \citep{bib:vainshtein,bib:nicolis,bib:brax11}.

Braneworld constructions and massive gravity models give rise to an explicit disformal coupling to matter of the form $\sim \partial_\mu \pi \partial_\nu \pi T^{\mu\nu}$ (see e.g. \cite{bib:trodden}). 
The disformal coupling plays a role in the field cosmological evolution, which makes  this kind of Galileon model interesting to compare with cosmological data. Such a coupling between a scalar field and matter has also been widely studied in \cite{bib:koivisto,bib:zumalacarregui,bib:brax12b,bib:brax13,bib:brax14}; and \cite{bib:brax15}. 

The uncoupled Galileon model has already been constrained by observational data in \citet{bib:ali}, \citet{bib:appleby2}, \citet{bib:okada}, \citet{bib:nesseris}, and more recently in \cite{bib:neveu} (hereafter \citetalias{bib:neveu}), \cite{bib:neveu14} (hereafter \citetalias{bib:neveu14}),  \cite{bib:barreira13}, \cite{bib:barreira14}, and \cite{bib:barreira14b}. In \citetalias{bib:neveu}, we introduced a new parametrisation of the model which allowed us to break degeneracies between Galileon parameters and to constrain them independently of initial conditions on the $\pi$ field. The same methodology was adopted here, and we refer the interested reader to \citetalias{bib:neveu} for more details. Moreover, in \citetalias{bib:neveu14}, we tested for the first time a Galileon model disformally coupled to matter and showed that a non-zero disformal coupling was preferred at the $2.5\sigma$ level by cosmological data, favouring a braneworld origin of the Galileon theory. In \citetalias{bib:neveu} and \citetalias{bib:neveu14}, we concluded that the uncoupled Galileon model provides as good an agreement with current data as the $\Lambda$CDM model. More recently, \cite{bib:barreira14} and \cite{bib:barreira14b} have shown that the $\Lambda$CDM  model was favoured over the uncoupled Galileon model  restricted to its tracker solution  because of tensions between cosmological data sets. To compare our conclusions, this particular case of the Galileon theory was also explored in the present paper.

In this paper, our aim was to set cosmological constraints on the Galileon conformal coupling to matter and to update the $\Lambda$CDM and Galileon constraints with the latest available cosmological data. In particular, for the cosmic microwave background (CMB), we used distance priors derived from the \Planck satellite polarised data \citep{bib:planck15}. The latest baryonic acoustic oscillations (BAO) measurements were  also included, as well as the latest growth of structure measurements. 

Section~\ref{sec:data} describes our updated data sets. Section~\ref{sec:standard} shows their impact on the cosmological standard model constraints. The Galileon theory is introduced in Section~\ref{sec:galileon} and the corresponding cosmological constraints are described in Section~\ref{sec:results}. Section~\ref{sec:disc} discusses these results and their implications, in particular when including non-cosmological data to constrain the Galileon couplings. We conclude in Section~\ref{sec:concl}.

\section{Data sets}\label{sec:data}

In this work, we follow the same methodology developed in \citetalias{bib:neveu}, with changes described below. 

\subsection{Type Ia supernovae}

We  used the recent type Ia supernovae (SNe Ia) sample published jointly by the SuperNova Legacy Survey (SNLS) and the Sloan Digital Sky Survey (SDSS) collaborations \citep{bib:betoule13,bib:jla}. This SN~Ia sample is  referred to as the Joint Light-curve Analysis (JLA) sample in the following. The 740 supernovae with their full systematic and statistic covariance matrices were considered. We recall that, usually, one should fit and marginalise over two nuisance parameters $\alpha$ and $\beta$, which describe the SN Ia variability in stretch and colour \citep{bib:astier,bib:guy2010,bib:conley}. However, in \citetalias{bib:neveu} it was shown that for the Galileon model we can keep $\alpha$ and $\beta$ fixed to their marginalised values as found in the $\Lambda$CDM model with the same data set. In this study, as in \citetalias{bib:neveu14}, we thus took  the $\alpha$ and $\beta$ values directly from \cite{bib:jla}.

\subsection{Updated CMB data}\label{sec:cmb}

In this work, as in \citetalias{bib:neveu} and \citetalias{bib:neveu14}, we did not compute a full CMB power spectrum in the frame of the Galileon theory. Instead, simple CMB priors were considered. These priors contain mostly information from the first acoustic peak of the CMB power spectrum, and thus are less constraining than the full CMB power spectrum. However, they are easier to handle when dealing with complex alternative cosmological models, and contain enough information to set constraints. More specifically, the priors concern three parameters, which are $z_*$ the redshift of the last scattering surface, $l_a$ the acoustic scale related to the comoving sound speed horizon $r_s(z_*),$ and $R$ the shift parameter related to the angular distance between us and the last scattering surface.

\begin{table}[htb]
\caption[]{CMB distance priors.}
\label{tab:cmb}
\begin{center}
\begin{tabular}{cccc} \hline \hline \\ [-1ex]
   & WMAP9 & \Planck TT & \Planck TT,TE,EE \\  [1ex] \hline \\ [-1ex]
$R$  & $1.725\pm0.018$ & $1.7489\pm0.0074$ & $1.7492\pm0.0049$ \\ [1ex]  \hline \\ [-1ex]
$l_a$  &  $302.40\pm0.69$ & $301.76\pm0.14$ & $301.787\pm0.089$\\ [1ex]  \hline \\ [-1ex]
$z_*$  & $1090.88\pm1.00$ & $1090.00\pm0.43$ & $1089.99\pm0.29$\\ [1ex]  \hline \\ [-1ex]
\end{tabular}
\tablefoot{WMAP9 priors are quoted from \cite{bib:wmap9}. Planck priors are derived using the 2015 \Planck likelihoods with either the TT spectrum only (Col. 2) or the TT, TE, and EE spectra (Col. 3) as detailed in Table 3 of \cite{bib:planck15}. Correlations between the parameters are not presented in the table for clarity but are taken into account in our analysis.}
\end{center}
\end{table}

The \Planck collaboration recently released new likelihoods including data  on the temperature and polarisation of the CMB from the full mission. As shown in Table 3 of \cite{bib:planck15}, using polarisation data can reduce cosmological parameter uncertainties by a factor of 2 compared to using temperature data alone. To derive our CMB priors from the \Planck data, we used the 2015 likelihoods corresponding to a flat $w$-CDM cosmology and the full polarisation data (TT, TE, and EE power spectra)\footnote{Likelihoods can be retrieved from the \Planck Legacy Archive \url{http://www.cosmos.esa.int/web/planck/pla}. Scripts to derive parameters from this likelihood come from CosmoMC \citep{bib:cosmomc,bib:cosmomc2} as detailled in \url{http://cosmologist.info/cosmomc/readme_python.html}.}. The obtained priors are
\begin{equation}
\langle \mathbf{V}_{CMB}\rangle=\left( \begin{array}{c}  \langle l_a \rangle \\ \langle R \rangle \\ \langle z_* \rangle \end{array} \right)
= \left( \begin{array}{c} 301.787 \pm 0.089 \\ 1.7492 \pm 0.0049 \\ 1089.99 \pm 0.29 \end{array} \right),
\end{equation}
with the corresponding inverse covariance matrix
\begin{equation}
 \mathbf{C}^{-1}_{CMB}=\left( \begin{array}{ccc} 162.48 & -1529.4 & 2.0688 \\ -1529.4 & 207232 & -2866.8 \\ 2.0688 & -2866.8 & 53.572 \end{array} \right).
\end{equation}

It should be noted that the WMAP papers \citep{bib:komatsu09,bib:komatsu11,bib:wmap9} used a similar recipe to derive their CMB priors. 
Furthermore, as pointed out in \citet{bib:nesseris}, the Galileon model fulfils the assumptions required in \citet{bib:komatsu09} to use these distance priors when testing a dark energy model, namely a Friedmann-Lema\^itre-Robertson-Walker (FLRW) Universe with the standard number of neutrinos and a dark energy background with negligible interactions with the primordial Universe. For this point, we a posteriori checked that in all our best fit scenarios the amount of dark energy at decoupling remains below 2\% of the total energy content of the Universe. Finally, it has been shown in many works that the above CMB priors are independent of the dark energy model used to derive them (see  \cite{bib:wang07,bib:mukherjee,bib:elgaroy,bib:cai}).

In Table~\ref{tab:cmb}, these CMB priors are compared to the previous WMAP9 priors \citep{bib:wmap9} that were used in \citetalias{bib:neveu14} and to \Planck priors derived from temperature data only. Compared to \citetalias{bib:neveu14}, uncertainties on the priors are approximately divided by a factor of 5, with central values compatible within uncertainties. The gain on the cosmological constraints is detailed in Section~\ref{sec:standard}. We checked that the priors are consistent with those already derived by \cite{bib:wang13} and \cite{bib:cai} using the 2013 data release from the \Planck collaboration. 

To use the CMB priors, we  followed again the recipe recommended by \cite{bib:komatsu09}. The key point of the recipe consists in minimising $\chi^2_{\rm CMB}$ over the reduced Hubble constant $h = H_0 / (100 \rm{\,km/s/Mpc}) $ (where $H_0$ is the Hubble constant value) and the baryon density today $\Omega_b^0h^2$. We used the \cite{bib:hu} fitting formula to compute $z_*$.

Owing to the controversy between direct measurements of $H_0$ and indirect constraints from BAO and CMB data (see e.g. \cite{bib:boss14,bib:planck,bib:planck15}), we did not use a Gaussian prior on $H_0$ to guide the minimisation procedure over $h$. However, we report the existing direct measurements of $H_0$ in Table~\ref{tab:h0} for comparison with our constraints in the different models we studied.

\begin{table}[htb]
\caption[]{$H_0$ direct measurements.}
\label{tab:h0}
\begin{center}
\begin{tabular}{cc} \hline \hline \\ [-1ex]
 $H_0$ (km/s/Mpc)  & Reference  \\  [1ex] \hline \\ [-1ex]
$73.8\pm2.4$  & \cite{bib:riess11} \\
$74.3 \pm 2.1$ & \cite{bib:freedman}\\
$72.5\pm2.5$ & \cite{bib:efstathiou}\\
$70.6\pm3.3$ & \cite{bib:efstathiou} \\
$73.2\pm1.7$  & \cite{bib:riess16} \\  [1ex]  \hline 
\end{tabular}
\tablefoot{In this paper, as in \citetalias{bib:neveu14}, none of these measurements is used as a Gaussian prior on $H_0$, contrary to \citetalias{bib:neveu}.}
\end{center}
\end{table}

\subsection{Updated BAO and Lyman-$\alpha$ measurements}

\begin{table*}[htb]
\caption[]{Updated BAO (top) and Lyman-$\alpha$ (bottom) measurements.}
\label{tab:bao}
\begin{center}
\begin{tabular}{ccccccc} \hline \hline \\ [-1.ex]
$z$  & $D_V\left(\frac{r_d^{\rm fid}}{r_d}\right)$ (Mpc) & $H\left(\frac{r_d}{r_d^{\rm fid}}\right)$ (km/s/Mpc) & $D_A\left(\frac{r_d^{\rm fid}}{r_d}\right)$ (Mpc) & $r$ & Survey & Reference \\ [1.8ex] \hline  \\ [-1ex]
0.106 & $456 \pm 20$ & - & - &- & 6dFGS & \cite{bib:beutler} \\
0.15 & $664 \pm 25$ &  - & - &- & SDSS MGS & \cite{bib:ross} \\
0.32 & $1264 \pm 25$ & - & - &- & BOSS LOWZ & \cite{bib:tojeiro} \\
0.44 & $1716 \pm 83$ & - & - &- & WiggleZ & \cite{bib:kazin} \\
0.57 & - & $96.8\pm 3.4$ & $1421 \pm 20$ & 0.539 & BOSS CMASS & \cite{bib:boss14} \\ 
0.6 & $2221 \pm 101$ & - & - &- & WiggleZ & \cite{bib:kazin} \\
0.73 & $2516 \pm 86$ & - & - &- & WiggleZ & \cite{bib:kazin} \\
 [1ex] \hline \\ [-1.ex]
2.34 & - & $222\pm 7$ & $1662 \pm 96$ & 0.43 & BOSS DR11 & \cite{bib:delubac} \\ 
2.36 & - & $223\pm 7$ & $1616 \pm 60$ & 0.39 & BOSS DR11 & \cite{bib:fontribera} \\ [1ex] \hline
\end{tabular}
\tablefoot{Parameter $r$ is the cross-correlation between  $H(z)$ and $D_A(z)$ measurements. The $r$ values for the Lyman-$\alpha$ measurements are taken from \cite{bib:aubourg}. WiggleZ $D_V(r_d^{\rm fid}/r_d)$ three measurements are correlated: the full inverse covariance matrix from \cite{bib:kazin} is used to compute our $\chi^2$ values, but not detailed here for brevity.}
\end{center}
\end{table*}

BAO distances provide information on the imprint on the distribution of galaxies of the comoving sound horizon at recombination. 
BAO surveys exploit this standard ruler to derive observables which map the Universe expansion. These can be $H(z)$ the Hubble parameter derived jointly with $D_A(z)$ the angular distance if the survey is wide enough, or an effective distance $D_V(z)$  \citep{bib:eisenstein05} given by
\begin{equation}
D_V(z)=\left[(1+z)^2D_A^2(z)\frac{cz}{H(z)}\right]^{1/3} .
\end{equation}
Generally, constraints on these observables are derived by comparing data to mock catalogues built with a fiducial cosmology which fixes $r_d = r_s(z_d)$, the comoving sound horizon at the baryon drag epoch redshift $z_d$. In order to provide measurements independent of this fiducial choice, the observables are usually expressed in terms of a ratio $(r_d^{\rm fid}/r_d)$. Then, to compare predictions to these measurements, one has to compute $H(z)$ and $D_A(z),$ but also $r_d$ and $r_d^{\rm fid}$. To compute the $r_d^{\rm fid}$ values, the fiducial cosmology proper to each measurement is used. In our code, the baryon drag epoch redshift $z_d$ is computed using the \cite{bib:eisenstein98} fitting formula. The validity of this approximate formula to compute the ratio $(r_d^{\rm fid}/r_d)$ has been checked in \cite{bib:mehta} and is further discussed in Appendix~\ref{sec:rd}.  
We note  that the BAO constraints are computed together with CMB priors described in Section~\ref{sec:cmb} as both probes share the computation of the comoving sound horizon $r_s$ (at redshifts $z_d$ and $z_*$) and depend on $\Omega_b^0h^2$ and $h$.

In \citetalias{bib:neveu} and \citetalias{bib:neveu14}, only three BAO measurements were used,  from \cite{bib:beutler,bib:padmanabhan,bib:boss12}. Since then, the last two measurements have been updated by the BOSS collaboration at redshift $z=0.32$ and $z=0.57$ \citep{bib:tojeiro,bib:boss14}, and three new measurements from the WiggleZ survey have been released in \cite{bib:blake12} and re-analysed in \cite{bib:kazin}. The compilation of the BAO measurements used in the present paper is presented in Table~\ref{tab:bao}. We note  that  the full inverse covariance matrices reported in these papers are used to compute our $\chi^2$ values. The BAO data set presented in \cite{bib:planck15} is similar to ours, but they preferred not to use WiggleZ BAO measurements as there is a small overlap with the BOSS survey. The correlation between BOSS and WiggleZ measurements has been recently evaluated in \cite{bib:beutler15} and appears to be small. We thus decided to use  both data sets.

The above measurements use galaxies to extract the BAO scale at redshifts $z\lesssim 1$. The BAO feature was also detected at redshift $z\approx 2.3$ in the flux-correlation function of the Lyman-$\alpha$  (hereafter  Ly$\alpha$) forest of high-redshift quasars \citep{bib:delubac} and in the cross-correlation of quasars with the Ly$\alpha$ forest absorption \citep{bib:fontribera}, see Table~\ref{tab:bao}.
 \cite{bib:delubac} notes that these two measurements are nearly uncorrelated and can be used together to build stronger cosmological constraints.

The Galileon best fit scenarios from \citetalias{bib:neveu} and \citetalias{bib:neveu14} revealed important deviations from $\Lambda$CDM mostly at redshifts $z\lesssim 3$. The precise mapping of the Universe expansion in this redshift range could be used to discriminate between the cosmological models. The Ly$\alpha$ distance measurements at redshift $z\approx 2.3$ allow  the mapping of  the Universe expansion history at intermediate redshifts between the anchor of the CMB measurement at $z\approx 1100$ and the local $z \lesssim 1$ distance measurements from SNe~Ia and BAO surveys. 

In their cosmological paper \cite{bib:planck15}, the \Planck Collaboration decided not to use the two Ly$\alpha$ measurements as they exhibit a $\approx 2\sigma$ tension with $\Lambda$CDM (see also \cite{bib:delubac}). This kind of measurement is less mature than galaxy measurements and may be affected by still unknown systematics. However, as our aim is to look beyond $\Lambda$CDM, it would be biased to reject these two measurements because of this tension. In this paper, we preferred to use them and check their impact on the $\Lambda$CDM and Galileon models.

\subsection{Updated growth of structure data}

\begin{table*}[htb]
\caption[]{Updated growth of structure data.}
\label{tab:gof}
\begin{center}
\begin{tabular}{cccccc} \hline \hline \\ [-1ex]
$z$ & $f\sigma_8(z)$ & $F(z)$ & $r$ & Survey & Reference \\  [1ex] \hline \\ [-1ex]
0.067 & $0.423\pm 0.055$ & - & - &  6dFGRS (a) & \cite{bib:beutler12} \\ 
0.15 & $0.53\pm 0.19$ & - & - & SDSS MGS (b) & \cite{bib:howlett} \\ 
0.32 & $0.384\pm 0.095$ & $0.327\pm0.030$ & 0.71 & BOSS LOWZ & \cite{bib:chuang} \\ 
0.44 & $0.413\pm 0.080$ & $0.482\pm0.049$ & 0.73 & WiggleZ & \cite{bib:blake12} \\
0.57 & $0.441\pm0.044$ & - & - & BOSS CMASS (c) & \cite{bib:samushia14} \\ 
0.6 & $0.390\pm 0.063$ & $0.650\pm0.053$ & 0.74 & WiggleZ & \cite{bib:blake12} \\
0.73 & $0.437\pm 0.072$ & $0.865\pm0.073$ & 0.85 & WiggleZ & \cite{bib:blake12} \\
0.8 & $0.47\pm 0.08$ & - & - & VIPERS (a) & \cite{bib:delatorre} \\ [1ex]  \hline \\ [-1ex]
\end{tabular}
\tablefoot{Parameter $r$ is the cross-correlation between the  $f\sigma_8(z)$ and $F(z)$ measurements. Correlations between the three WiggleZ measurements also exist but are not reported here. (a) The Alcock-Paczynski effect is supposed to be negligible in these analysis. (b) Values of $f\sigma_8$ are corrected for the Alcock-Paczynski effect but no $F(z)$ values are provided. (c) \cite{bib:samushia14} provides a $F(z)$ measurement, but to avoid double counting with the BOSS CMASS $H(z)$ and $D_A(z)$ measurements in Table~\ref{tab:bao} here we used only their $f\sigma_8(z)$ result marginalised over $F(z)$.}
\end{center}
\end{table*}

In \citetalias{bib:neveu} and \citetalias{bib:neveu14}, we used nine linear growth rate measurements $f\sigma_8(z)$ jointly with five Alcock-Paczynski parameter $F(z)$ measurements to compute constraints on the Universe growth history. We chose measurements that do not use an underlying fiducial $\Lambda$CDM cosmology but rely on the Alcock-Paczynski test  \citep{bib:alcock} to derive a value for $f\sigma_8$ from raw data.  In the above, $f(z)$ stands for the linear growth rate of structures and $\sigma_8(z)$ for the RMS of matter fluctuations in spheres of $8h^{-1}\mathrm{Mpc}$ radius. The Alcock-Paczynski parameter is defined as
\begin{equation}
F(z) = \frac{1+z}{c}D_A(z) H(z).
\end{equation}

The precise methodology for  computing the above observables and the corresponding $\chi^2$ values is described in \citetalias{bib:neveu} and \citetalias{bib:neveu14}. In summary, $f(z)$ is computed by integrating the perturbation equations corresponding to the cosmological model under study. As regards $\sigma_8(z)$, since we are dealing with the linear growth of structures, we followed the \cite{bib:samushia12} method
\begin{equation}\label{eq:s81}
\sigma_8(z) = \sigma_8(z_*)\frac{D(z)}{D(z_*)},
\end{equation}
with
\begin{equation}\label{eq:s82}
\sigma_8(z_*) = \sigma_8^0\frac{D^{\Lambda\mathrm{CDM}}(z_{*})}{D^{\Lambda\mathrm{CDM}}(0)},
\end{equation}
where $D(z)$ is the linear growth of structures and $\sigma_8^0$ the present value of the power spectrum normalisation derived from CMB data in the $\Lambda$CDM model. Doing so we assumed that the $\sigma_8$ value at decoupling is the same in all cosmological models, which is the case if dark energy was subdominant during the radiation era. This hypothesis is consistent with the assumptions made to derive the CMB priors in section~\ref{sec:cmb}.

In this paper we use the updated growth measurements  presented in Table~\ref{tab:gof}, together with their full covariance matrix. We also updated the present value of $\sigma_8$ to
\begin{equation}
\sigma_8^0 = 0.8150 \pm 0.0087
\end{equation}
from \cite{bib:planck15} using CMB polarisation information. This value is $1\sigma$ lower than that used in \citetalias{bib:neveu14}, $\sigma_8^0 = 0.829 \pm 0.012$ from \cite{bib:planck}. It is used as a common seed at $z=10^3$ (after removing growth evolution from $z=0$ to $z=10^3$, see equations~\ref{eq:s81} and~\ref{eq:s82}) for the standard and Galileon models. 

Growth of structure measurements are a key probe for  testing modified gravity models. However, they are also the most difficult to handle as only the linear growth rate of perturbations is easily computable in general relativity and modified gravity theories. Non-linearities play an important role at small scales, but their inclusion requires heavy simulations, in particular for the modified gravity theories which include screening mechanism through non-linear features. 

It is thus important to recall   the main hypothesis  when testing a Galileon theory with growth data. In our analysis, we used linear growth rate measurements which include scales up to $\approx 0.1\,h\mathrm{Mpc}^{-1}$ measured through redshift space distortions (RSD). Most measurements are derived from raw data using models that encompass non-linearities which are thought to be important at scales above $0.05\,h\mathrm{Mpc}^{-1}$ \citep{bib:jennings11,bib:blake11a}. The choice of the cut-off scale is an experimental compromise between cutting small scales in matter power spectra and increasing the statistical size of the galaxy samples. 

Our code includes only linear growth rate predictions based on linear theory to describe the growth of  matter and of Galileon field perturbations. In particular, this means that we assumed that the Galileon screening mechanism, which is due to the non-linear Galileon field Lagrangians, can be neglected over the range of scales probed by the observations. However, the transition scale between the screened and unscreened regimes is still difficult to estimate, but progress has been made. Heavy numerical simulations have been performed in the Galileon to characterise the impact of the non-linear screening mechanism on several
observables \citep{bib:li13,bib:barreira13b,bib:barreira13c,bib:gronke,bib:barreira14c,bib:barreira16}. \cite{bib:barreira14c} have shown that the screening sets  approximately on scales $k> 0.1\,h\mathrm{Mpc}^{-1}$ in the cubic ($c_4=c_5=0$) and quartic ($c_5=0$) Galileon models, but not for the full Galileon model. The corresponding threshold in the full Galileon model is difficult to estimate. However, with the above value it seems that present measurements lay at the frontier of the Vainshtein regime. Until future works indicate whether non-linear scales have a significant impact on present growth measurements, we choose to show how growth data constrain the Galileon model. 

Concerning the BOSS CMASS results at $z=0.57$, two measurements exist, \cite{bib:samushia14} and \cite{bib:beutler14}, derived using two different techniques. \cite{bib:beutler14} have shown that their measurement is less stable with respect to the values of the cut-off scale than the measurement from \cite{bib:samushia14}. In the light of the above discussion, we only used the latter. 
However, as the BOSS CMASS measurements of $H(z)$ and $D_A(z)$ are also included in our data (see Table~\ref{tab:bao}), we marginalised their value of $f\sigma_8(z)$ over $F(z)$ to avoid double counting.
Doing so, we keep the statistical power of the precise anisotropic BAO measurement of \cite{bib:boss14} instead of using the full RSD information in \cite{bib:samushia14}.\\

In summary, this paper encompasses seven more measurements than \citetalias{bib:neveu} and \citetalias{bib:neveu14} (principally in the BAO data set). We thus expect an increase in the total $\chi^2$ values by a similar amount.

\section{Cosmological constraints on standard cosmological models}\label{sec:standard}

To derive constraints on cosmological models (standard ones or the Galileon model), we sampled the model parameter space as in \citetalias{bib:neveu}, namely using a grid technique with a fixed step size to ensure that all parameter sets of interest are explored.   

\subsection{Cosmological constant model $\Lambda$CDM }\label{sec:lcdm}

\begin{figure}[hbtp]
\begin{center}
\epsfig{figure=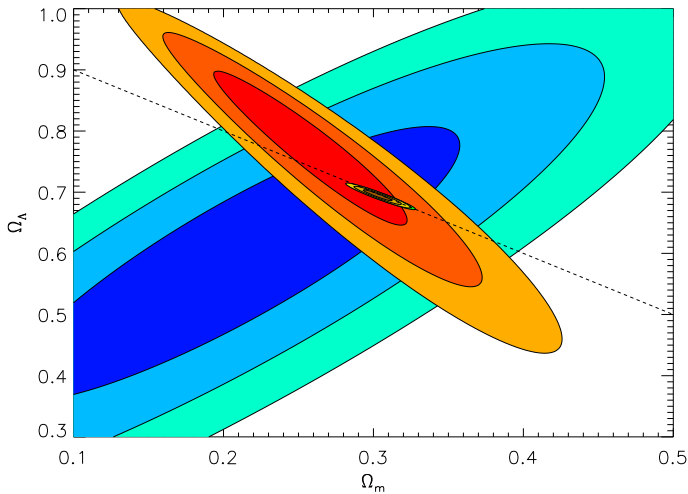, width=\columnwidth} 
\caption{Cosmological constraints on the $\Lambda$CDM model from JLA SNe~Ia (blue), growth data (red), \textit{Planck}+BAO+Ly$\alpha$ data (green, nearly masked by the yellow contour), and all data combined (yellow). The black dashed line indicates the flatness condition $\Omega_m+\Omega_\Lambda=1$.} 
\label{fig:lcdm}
\end{center}
\end{figure}

\begin{table*}
\caption[]{$\Lambda$CDM best fit values from different data samples}
\label{tab:lcdm}
\begin{center}
\begin{tabular}{ccccccc} \hline \hline \\ [-1ex]
Probe & $\Omega_m^0$ & $\Omega_\Lambda^0$ & $h$ & $\Omega_b^0h^2$ & $ \chi^2$ & $N_{\rm data}$ \\  [1ex] \hline \\ [-1ex]
SNe~Ia & $0.214^{+0.109}_{-0.103}$ &  $0.588^{+0.158}_{-0.157}$ & - & - & 691.0 & 740 \\ [1ex]  \hline \\ [-1ex] 
Growth & $ 0.265^{+0.048}_{-0.039}$ & $0.759^{+0.078}_{-0.091}$ & - & - & 2.9  & 12 \\ [1ex]  \hline \\ [-1ex]
\textit{Planck}+BAO+Ly$\alpha$ & $0.305^{+0.007}_{-0.006}$ & $0.693^{+0.006}_{-0.006}$ & 0.695 & 0.0240 & 14.5 & 15 \\ [1ex]  \hline \\ [-1ex]
All & $0.303^{+0.007}_{-0.006}$ & $0.695^{+0.006}_{-0.006}$ & 0.697 & 0.0241 & 710.6  & 767 \\ [1ex]  \hline \hline \\ [-1ex]
All \citetalias{bib:neveu14} & $0.284^{+0.012}_{-0.009}$ & $0.720^{+0.015}_{-0.012}$ & 0.689 & 0.0226 & 705.5 & 760 \\ [1ex]  \hline
\end{tabular}
\tablefoot{The JLA SNe~Ia is used with systematics included; $\alpha$ and $\beta$ are fixed to their marginalised value. $h$ and $\Omega_b^0h^2$ have been minimised so no error bars are provided. $N_{\rm data}$ is the number of measurements in each data set.}
\end{center}
\end{table*}

We used all the data presented in Section~\ref{sec:data} to constrain the cosmological constant $\Lambda$CDM model. The results are presented in Figure~\ref{fig:lcdm} and Table~\ref{tab:lcdm}. 

The use of the latest \Planck priors combined with the latest BAO measurements is responsible for a reduction of a factor of 2 of the uncertainties on the $\Lambda$CDM parameters, compared to what we observed with the previous data set. The $\Omega_m^0$ best fit value increased from $\approx0.28$ to $0.30$, as expected when using \Planck data instead of WMAP9 priors \citep{bib:planck15}. 

Tests have been conducted with and without the Lyman-$\alpha$ measurements to see the evolution of $\chi^2$. Without them, the same best fit as that in the third row of Table~\ref{tab:lcdm} is obtained, but the $\chi^2$ decreased from 14.5 to 4.5. This reflects the tensions reported for these measurements with the $\Lambda$CDM model \citep{bib:delubac,bib:aubourg}. 

Compared to \citetalias{bib:neveu}, there is a $1\sigma$ shift in the best fit values from growth data, leading to better agreement with $\Lambda$CDM. Two factors explain that change. Half of the shift is due to the updated $f\sigma_8$ measurements and the other half comes from the lower $\sigma_8^0$ value. A lower $\sigma_8^0$ value is indeed favourable to a universe with less matter and more dark energy in general as $\sigma_8^0$ describes how matter is distributed in $8\,h^{-1}\mathrm{Mpc}$ spheres.

The minimised $h$ values in Table~\ref{tab:lcdm} are compatible with other cosmological studies from e.g. \cite{bib:planck15} and \cite{bib:boss14}, but $\Omega_b^0h^2$ is higher than in these studies. This is probably due to the use of the approximate \cite{bib:eisenstein98} formula in our code for $z_*$.
The tension with the $H_0$ \cite{bib:riess11} measurement presented in Table~\ref{tab:h0} is still present.

Compared to \citetalias{bib:neveu} and \citetalias{bib:neveu14}, the agreement between all probes is better since the final $\chi^2$ moved from 705.5 in \citetalias{bib:neveu14} to 710.6 while adding seven new measurements. 
Our global best fit is compatible with the \cite{bib:planck15} $\Lambda$CDM best fit using all data (last column in their Table 4) and uncertainties are of the same size despite the use of CMB derived parameters instead of the full power spectrum.  

 \subsection{FWCDM model}\label{sec:fwcdm}

\begin{figure}[hbtp]
\begin{center}
\epsfig{figure=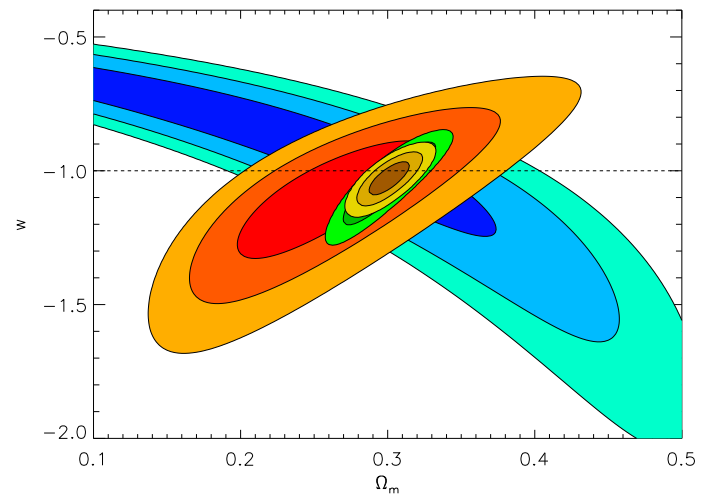, width=\columnwidth} 
\caption{Cosmological constraints on the FWCDM model from SNe~Ia (blue), growth data (red), \textit{Planck}+BAO+Ly$\alpha$ data (green), and all data combined (yellow).} 
\label{fig:fwcdm}
\end{center}
\end{figure}

\begin{table*}[htb]
\caption[]{FWCDM best fit values from different data samples}
\label{tab:fwcdm}
\begin{center}
\begin{tabular}{ccccccc} \hline \hline \\ [-1ex]
Probe & $\Omega_m^0$ & $w$ & $h$ & $\Omega_b^0h^2$ & $ \chi^2$ & $N_{\rm data}$ \\  [1ex] \hline \\ [-1ex]
SNe~Ia & $0.231^{+0.112}_{-0.132}$ &  $-0.92^{+0.20 }_{-0.23}$ & - & - & 691.7 & 740\\ [1ex]  \hline \\ [-1ex] 
Growth & $0.261^{+0.048}_{-0.039}$ & $-1.11^{+0.14}_{-0.15}$ & - & - & 3.0 & 12 \\ [1ex]  \hline \\ [-1ex]
\textit{Planck}+BAO+Ly$\alpha$  & $0.301^{+0.013}_{-0.012}$ & $-1.04^{+0.06}_{-0.06}$ & $0.698$ & $0.0241$ & 15.5 & 15 \\ [1ex]  \hline \\ [-1ex]
All  & $0.301^{+0.010}_{-0.008}$ & $-1.03^{+0.04}_{-0.04}$ & $0.697$ & $0.0241$ & 711.7 & 767 \\ [1ex]  \hline \hline \\ [-1ex]
All \citetalias{bib:neveu14} & $0.294^{+0.014}_{-0.010}$ & $-0.93^{+0.05}_{-0.04}$ & 0.678 & 0.0226 & 703.3 & 760 \\ [1ex]  \hline
 \end{tabular}
\tablefoot{The JLA SNe~Ia is used with systematics included; $\alpha$ and $\beta$ are fixed to their marginalised value. $h$ and $\Omega_b^0h^2$ have been minimised so no error bars are provided. $N_{\rm data}$ is the number of measurements in each data set.}
\end{center}
\end{table*}

All data sets presented in Section~\ref{sec:data} were used to constrain another standard cosmological model, FWCDM, which assumes a constant equation of state $w$ for dark energy and a flat Universe. Results are presented in Figure~\ref{fig:fwcdm} and Table~\ref{tab:fwcdm}. The combination of all data sets leads to a $w$ best fit value compatible with $-1$ and the final $\chi^2$ shifts from 703.3 to 711.7 compared to \citetalias{bib:neveu14}. The tendencies observed in Section~\ref{sec:lcdm} are also present in this model. The total $\chi^2$ is equivalent to the $\Lambda$CDM value.

\section{Galileon theory}\label{sec:galileon}

\subsection{Lagrangians}\label{sec:lagrangians}

In this paper, as in \citetalias{bib:neveu} and \citetalias{bib:neveu14}, we use the Galileon covariant action with the parametrisation of~\cite{bib:appleby} established in the Jordan frame,

\begin{equation}\label{eq:action}
S=\int d^4x \sqrt{-g}\left( \frac{M_P^2R}{2} - L_m - \frac{1}{2} \sum_{i=1}^{5} c_i L_i -  L_G -  L_0\right),
\end{equation}
with $M_P$ the Planck mass, $R$ the Ricci scalar, $g$ the determinant of the metric $g_{\mu\nu}$, $L_m$ the matter Lagrangian, and $L_i$ the Galileon Lagrangians. 
The $c_i$s are the arbitrary dimensionless parameters of the Galileon model that weight the different terms. The Galileon Lagrangians have a covariant formulation derived in \cite{bib:deffayet}

\begin{align}
L_1&=M^3\pi,\, L_2=(\pi_{;\mu} \pi^{;\mu}),\,
L_3=(\pi_{;\mu} \pi^{;\mu})(\square \pi)/M^3, \notag \\
L_4&=(\pi_{;\mu} \pi^{;\mu})\left[ 2(\square \pi)^2 - 2 \pi_{;\mu\nu}\pi^{;\mu\nu} - R(\pi_{;\mu} \pi^{;\mu})/2 \right]/M^6, \notag  \\
\begin{split}
L_5&=(\pi_{;\mu} \pi^{;\mu})
\left[ (\square \pi)^3 - 3(\square \pi) \pi_{;\mu\nu}\pi^{;\mu\nu}
+2\pi_{;\mu}^{\ ;\nu}\pi_{;\nu}^{\ ;\rho}\pi_{;\rho}^{\ ;\mu}\right.\\&\ \ \left. -6 \pi_{;\mu}\pi^{;\mu\nu}\pi^{;\rho}G_{\nu\rho}\right]/M^9,
\end{split} \notag \\
L_G&=c_G M_P G^{\mu\nu}\pi_{;\mu} \pi_{;\nu}/M^3,\, L_0 = c_0 M_P R \pi
\end{align}
\noindent where $\square \pi \equiv \pi^{;\mu}_{\ ;\mu}$,  $M$ is a mass parameter defined as $M^3=H_0^2M_P$.  With this definition the $c_i$ free parameters are dimensionless.

$L_2$ is the usual kinetic term for a scalar field, while $L_3$ to $L_5$ are non-linear couplings of the Galileon field to itself, to the Ricci scalar $R$, and to the Einstein tensor $G_{\mu\nu}$, providing the necessary features for modifying gravity and mimicking dark energy. $L_1$ is a tadpole term that acts as the usual cosmological constant, and may furthermore lead to vacuum instability because it is an unbounded potential term. Therefore, in the following we set $c_1=0$.

$L_G$ and $L_0$ are Lagrangians appearing in the generalised Galileon theory. They can be understood, correspondingly, as direct disformal and conformal couplings between matter and the Galileon field when translated in the Einstein frame and in the weak field limit \citep{bib:appleby}. They introduce two dimensionless and constant parameters $c_G$ and $c_0$. Defining $T^{\mu\nu}$ as the matter energy-momentum tensor, in the Einstein frame these two Lagrangians are
\begin{equation}\label{eq:couplings}
L_G=\frac{c_G}{M_P M^3}\partial_\mu \pi \partial_\nu \pi T^{\mu\nu},\, L_0=\frac{c_0}{M_P}\pi T^{\mu}_{\ \mu}.
\end{equation}
Given these expressions, $L_G$ is called the disformal coupling to matter and $L_0$ the conformal coupling to matter. Although throughout this paper we work in the Jordan (physical) frame, where definitions from equation~\ref{eq:couplings} are not present, we loosely use the terms ``disformal'' and ``conformal'' couplings when referring to $c_G$ and $c_0$. With this convention by uncoupled Galileon we mean $c_G=c_0=0$. The disformal coupling is motivated by extra-dimension considerations. It naturally arises in the decoupling limit of massive gravity (see \citealt{bib:deRhamMasGra2}). It also automatically arises when dealing with a fluctuating 3+1 brane in a $D=4+n$ dimensional bulk when matter lives exclusively in the brane. This disformal coupling has already been studied in scalar field theories as reported in \cite{bib:brax12b,bib:brax13}, and \cite{bib:brax14}. The disformal coupling to photons can play a role in gravitational lensing and has an observational signature (see \citealt{bib:wyman}).
In the more general context of scalar field theories, the disformal coupling has been recently constrained in particle physics using Large Hadron Collider data \citep{bib:monophoton,bib:brax15}.

The action in equation~\ref{eq:action} leads to three differential equations: two Einstein equations (with (00) the temporal component and (ij) the spatial one) coming from the action variation with respect to the metric $g_{\mu\nu}$, and the scalar field equation of motion, from the action variation with respect to the $\pi$ field. The equations are given explicitly in Appendix B of \cite{bib:appleby}. With these three differential equations the evolution of the Universe and the dynamics of the field can be computed. 

To solve the cosmological equations, we chose the Friedmann-Lema\^itre-Robertson-Walker (FLRW) metric. The functions to compute are the Hubble parameter $H=\dot a /a$ (with $a$ the cosmic scale factor), $y=\pi/M_P$, and $x=\pi'/M_P$, with a prime denoting $d/d \ln a$ (see \cite{bib:appleby} and Sect.~\ref{sec:eq}).

The rest of this section describes the Galileon equations with couplings that can be compared with the uncoupled case detailed in \citetalias{bib:neveu}.

\subsection{Initial conditions}\label{sec:inicond}

\par To compute the solutions of these cosmological equations, we need to set initial conditions, in particular for $x$ as  was shown in  \citetalias{bib:neveu}.
We arbitrarily chose to define this initial condition at $z=0$, which we denote $x_0=x(z=0)$. There is no prior information about the value of the Galileon field or its derivative at any epoch. But, as shown in \citetalias{bib:neveu}, $x_0$ can be absorbed by redefining the $c_i$s and functions as 
\begin{align}
\bar  c_i &= c_i x_0^i \text{ for } i=2...5,\ \bar c_G = c_G x_0^2,\ \bar c_0  = c_0 x_0,  \\
\x &= x/x_0,\ \y = y/x_0,\ \h = H/H_0 .
\end{align}
In several papers \citep{bib:appleby2,bib:nesseris,bib:neveu,bib:barreira13}, a degeneracy between the $c_i$s and $x_0$ was noted which prevents the Galileon model from being compared with data. Our parametrisation avoids this problem by absorbing the degeneracy between the $c_i$s and $x_0$ into the $\bar c_i$s (more details in \citetalias{bib:neveu}). This redefinition allows us to avoid treating the initial condition $x_0$ as an extra free parameter of the model. Doing so, the $\bar c_i$s remain dimensionless, and the initial conditions are simple for $\x$ and $\h$:
\begin{equation}\label{eq:initial}
\x_0 = 1,\quad \h_0 = 1 .
\end{equation}
The initial conditions for $\bar y$ are  discussed in the next section.

\subsection{Cosmological equations}\label{sec:eq}

\par To compute cosmic evolution in the Galileon model, we assume for simplicity that the Universe is spatially flat, in agreement with current observations. We used the FLRW metric in a flat space
\begin{equation}
ds^2=-dt^2+a^2\delta_{ij}dx^idx^j .
\end{equation}
When writing the cosmological equations, we can mix the (ij) Einstein equation and the $\pi$ equation of motion to obtain the following system of differential equations for $\x$, $\y$, and $\h$

\begin{align}
\y' &= \x, \\
\x' &=-\x + \frac{\alpha\lambda - \sigma\gamma}{\sigma\beta - \alpha\omega}, \label{eq:dpi}\\
\h'  &= \frac{\omega\gamma - \lambda\beta}{\sigma\beta - \alpha\omega}, \label{eq:dh}
\end{align}
with
\begin{align}
\begin{split}
\alpha &= \frac{\bar  c_2}{6}\h \x -3\bar  c_3\h^3\x^2 + 15\bar  c_4\h^5\x^3  -\frac{35}{2}\bar  c_5\h^7\x^4 \\ &\ \ + \bar c_0 \h -3\bar c_G \h^3 \x, 
\end{split}\notag \\
\gamma &= \frac{\bar  c_2}{3}\h^2\x - \bar  c_3\h^4\x^2  + \frac{5}{2}\bar  c_5\h^8\x^4 +2 \bar c_0 \h^2 - 2 \bar c_G \h^4 \x, \notag \\
\beta &= \frac{\bar  c_2}{6}\h^2 -2\bar  c_3\h^4\x + 9\bar  c_4\h^6\x^2 - 10\bar  c_5\h^8\x^3 - \bar c_G \h^4, \notag \\
\begin{split}
\sigma &= 2(1-2\bar c_0 \bar y)\h - 2\bar c_0 \h \x + 2\bar  c_3\h^3\x^3 - 15\bar  c_4\h^5\x^4 \\ &\ \ + 21\bar  c_5\h^7\x^5 +6 \bar c_G \h^3 \x^2, 
\end{split} \notag \\
\omega &= 2\bar  c_3\h^4\x^2 - 12\bar  c_4\h^6\x^3 + 15\bar c_5 \h^8\x^4 -2 \bar c_0 \h^2 + 4 \bar c_G \h^4 \x, \notag  \\
\begin{split}
\lambda &= 3(1-2\bar c_0 \bar y)\h^2- 2\bar c_0 \h^2 \x + \frac{\Omega_r^0}{a^4} +\frac{\bar c_2}{2}\h^2\x^2 - 2\bar  c_3\h^4\x^3 \\ &\ \ + \frac{15}{2}\bar c_4\h^6\x^4  - 9\bar  c_5 \h^8\x^5 - \bar c_G \h^4 \x^2, \label{eq:lambda}
\end{split}
\end{align}
as derived in the formalism of \cite{bib:appleby}, but using our normalisation for the $c_i$s. We obtain the same equations except that the $c_i$s are changed to $\bar c_i$s, and that we have a different treatment for the initial conditions. Equations~\ref{eq:dpi} and \ref{eq:dh} depend only on the $\bar c_i$s and $\Omega_r^0$. The radiation energy density in equation \ref{eq:lambda} is computed from the usual formula $\Omega_r^0=\Omega_\gamma^0(1+0.2271 N_{\mathrm{eff}})$ with $N_{\mathrm{eff}}=3.04$ the standard effective number of neutrino species \citep{bib:mangano}.  The photon energy density at the current epoch is given by $\Omega_\gamma^0h^2=2.469\times 10^{-5}$ for $T_{\mathrm{CMB}} =2.725\ \mathrm{K}$. 

If $\bar c_0 = 0$ whatever the value of $\bar c_G$, the differential equation system is only of first order and can be solved with the two initial conditions in equation~\ref{eq:initial}. If $\bar c_0\neq 0$, the differential equation system becomes of second order. Thus, an initial condition $\y_0$ must be set. From the (00) Einstein equation
\begin{equation}\label{eq:00}
\begin{split}
(1-2& \bar c_0 \y)\h^2 =\frac{\Omega_m^0}{a^3}+\frac{\Omega_r^0}{a^4} + \frac{\bar  c_2}{6}\h^2\x^2  - 2\bar  c_3\h^4\x^3 
\\&\ \   + \frac{15}{2}\bar c_4\h^6\x^4 - 7 \bar c_5\h^8\x^5-3\bar c_G \h^4\x^2 + 2 \bar c_0 \h^2 \x,
\end{split}
\end{equation}
it can be noted that the effect of $\y$ in the Friedmann equation is to renormalise the Newton constant $G_N$  by a factor $(1-2\bar c_0 \y)$ \citep{bib:brax15}. In order to avoid adding a new parameter $y_0$ to constrain, it is physically motivated to restrain our study to the case where the Newton constant at cosmological scales has its standard value today, i.e. $\y_0=0$.

\subsection{Perturbation equations}\label{sec:pert}

\par To test the Galileon model predictions for the growth of structures, we also need the equations describing density perturbations. We followed the approach of \citet{bib:appleby} for the scalar perturbation. \citet{bib:appleby} performed their computation in the frame of the Newtonian gauge, for scalar modes in the subhorizon limit, with the following perturbed metric:

\begin{equation}
ds^2=-(1+2\psi)dt^2+a^2(1-2\phi)\delta_{ij}dx^idx^j.
\end{equation}

In this context, the perturbed equations of the (00)~Einstein equation, the (ij) Einstein equation, the $\pi$ equation of motion, and the equation of state of matter are, in the quasi-static approximation (proved to be valid in the Galileon model by \cite{bib:barreira})

\begin{align}
\frac{1}{2} \kappa_4 \bar \nabla^2 \psi - \kappa_3\bar \nabla^2 \phi &= \kappa_1\bar \nabla^2 \delta y, \label{eq:perturbed1}\\
\kappa_5 \bar \nabla^2 \delta y - \kappa_4 \bar \nabla^2 \phi &= \frac{a^2 \rho_m}{H_0^2 M_P^2}\delta_m, \\
\frac{1}{2} \kappa_5 \bar \nabla^2 \psi - \kappa_1\bar \nabla^2 \phi &= \kappa_6\bar \nabla^2 \delta y, \\
\h^2\delta_m''+\h\h'\delta_m'+&2\h^2 \delta_m'=\frac{1}{a^2}\bar \nabla^2 \psi, \label{eq:perturbed2}
\end{align}
where $\delta y = \delta \pi / M_P$ is the perturbed Galileon, $\bar \nabla = \nabla / H_0$, $\rho_m$ is the matter density, and $\delta_m=\delta \rho_m / \rho_m$ is the matter density contrast. The formula for $\kappa_i$s are the same as in \citet{bib:appleby}, but rewritten following our parametrisation
\begin{align}
\begin{split}
\kappa_1 &= -6\bar c_4 \h^3\x^3\left(\h'\x+\h\x'+\frac{\h\x}{3}\right)\\ &\ \ +\bar c_5\h^5\x^3(12\h\x' + 15\h'\x + 3\h\x)  \\ &\ \ +2\bar c_G (\h\h'\x + \h^2\x'+\h^2\x)  -2 \bar c_0  , 
\end{split}\notag \\
\begin{split}
\kappa_3 &= -(1-2\bar c_0 \y) -\frac{\bar c_4}{2}\h^4\x^4 - 3\bar c_5 \h^5\x^4(\h'\x+\h\x')\\&\ \ +\bar c_G \h^2\x^2 ,
\end{split}\notag \\
\kappa_4 &= -2(1-2\bar c_0 \y) + 3 \bar c_4 \h^4\x^4 - 6 \bar c_5 \h^6\x^5 - 2\bar c_G \h^2\x^2, \notag \\
\kappa_5 &= 2 \bar c_3 \h^2\x^2 - 12\bar c_4 \h^4 \x^3 + 15 \bar c_5 \h^6\x^5+4\bar c_G\h^2\x -2\bar c_0,\notag\\
\begin{split}
\kappa_6 &= \frac{\bar c_2}{2} - 2\bar c_3(\h^2\x' + \h\h'\x + 2\h^2\x) -\bar c_G(2\h\h'+3\h^2) \\ &\ \ + \bar c_4 (12 \h^4\x\x' + 18 \h^3\x^2\h' + 13 \h^4\x^2) \\ &\ \ - \bar c_5(18\h^6\x^2\x'+30 \h^5\x^3\h' + 12 \h^6\x^3) .
\end{split}
\end{align}

With equations \ref{eq:perturbed1} to \ref{eq:perturbed2}, we can obtain a Poisson equation for $\psi$ with an effective gravitational coupling $G_{\mathrm{eff}}^{(\psi)}$ that varies with time and depends on the Galileon model parameters $\bar c_i$s:
\begin{equation}
\bar \nabla^2 \psi = \frac{4\pi a^2 G_{\mathrm{eff}}^{(\psi)}\rho_m}{H_0^2}\delta_m,
\end{equation}
\begin{equation}\label{eq:geff}
G_{\mathrm{eff}}^{(\psi)}=\frac{4(\kappa_3 \kappa_6 - \kappa_1^2) }{\kappa_5(\kappa_4 \kappa_1 - \kappa_5 \kappa_3) - \kappa_4(\kappa_4 \kappa_6 - \kappa_5 \kappa_1)} G_N.
\end{equation}
These equations can be used to compute the growth of matter perturbations in the frame of the Galileon model. Tensorial perturbation modes also exist, and are studied in Sect.~\ref{sec:tensorial}.

\subsection{Theoretical constraints}\label{sec:theoconstraints}

\par With at least six free parameters ($\Omega_m^0, h$, and the various $\bar c_i$s), it is necessary to restrict the parameter space on theoretical grounds before comparing the model to data. The theoretical constraints arise from multiple considerations: using the (00)~Einstein equation, requiring positive energy densities, and avoiding instabilities in scalar and tensorial perturbations.

\subsubsection{The (00)~Einstein equation and $\bar c_5$}

\par Because we used only the (ij) Einstein equation and the $\pi$ equation of motion to compute the dynamics of the Universe (equations \ref{eq:dpi} and \ref{eq:dh}), we are able to use the (00)~Einstein equation as a constraint on the model parameters.
More precisely, we used this constraint both at $z=0$ to fix one of our parameters and at other redshifts to check the reliability of our numerical computations. The parameter we chose to fix at $z=0$ is
\begin{equation}\label{eq:c5}
\bar c_5 = \frac{1}{7}\left(-1+\Omega_m^0 + \Omega_r^0 + \frac{\bar c_2}{6} - 2 \bar c_3 + \frac{15}{2} \bar c_4 + 2\bar c_0 - 3\bar c_G\right) .
\end{equation}
We chose to fix $\bar c_5$ as a function of the other parameters because allowing it to float introduces significant numerical difficulties when solving equations~\ref{eq:dpi} and \ref{eq:dh} since it represents the weight of the most non-linear term in these equations. As $\Omega_r^0$ is fixed given $h$, our parameter space has been reduced to $\Omega_m^0, h, \bar c_2,\bar c_3$, $\bar c_4$ and the optional Galileon couplings to matter.

\subsubsection{Positive energy density}

\par We require that the energy density of the Galileon field be positive from $z=0$ to $z=10^7$. At every redshift in this range, this constraint amounts to
\begin{align}\label{eq:rhopi}
\frac{\rho_{\pi}}{H_0^2 M_P^2} & =\frac{\bar c_2}{2}\h^2\x^2  - 6\bar c_3\h^4\x^3 + \frac{45}{2}\bar c_4\h^6\x^4 - 21 \bar c_5\h^8\x^5 \notag \\
&\ \ -9\bar c_G \h^4\x^2 + 6\bar c_0(\h^2 \x + \h^2 \y)> 0 .
\end{align}
This constraint is not really necessary for generic scalar field models and has actually no impact on the parameter space  because the other theoretical conditions described below are stronger (see \citetalias{bib:neveu}). We kept it for consistency with previous works.

\begin{table*}[htb]
\caption[]{Uncoupled Galileon model best fit values from different data samples. The combination of all distance measurements from  JLA+\textit{Planck}+BAO+Ly$\alpha$ data is denoted JPBL in the following.}
\label{tab:results}
\begin{center}
\begin{tabular}{ccccccccc} \hline \hline \\ [-1ex]
Probe & $\Omega_m^0$ & $\bar c_2$ & $\bar c_3$ & $\bar c_4$ & $h$ & $\Omega_b^0h^2$ & $ \chi^2$  & $N_{\rm data}$ \\  [1ex] \hline \\ [-1ex]
SNe~Ia & $0.328^{+0.055}_{-0.047}$ & $-4.2^{+1.7}_{-3.0}$ & $-1.3^{+1.0}_{-1.5}$ & $-0.48^{+0.46}_{-0.35}$ & - & - & 692.8 & 740 \\ [1ex]  \hline \\ [-1ex] 
Growth & $0.206^{+0.053}_{-0.043}$ & $-5.7^{+1.2}_{-2.0}$ & $-1.9^{+0.6}_{-1.2}$ & $-0.64^{+0.35}_{-0.26}$ & - & - & 2.9 & 12 \\ [1ex]  \hline \\ [-1ex] 
\textit{Planck}+BAO+Ly$\alpha$ & $0.279^{+0.008}_{-0.007}$ & $-5.4^{+1.9}_{-2.8}$ & $-1.9^{+0.9}_{-1.4}$ & $-0.63^{+0.46}_{-0.31}$ & 0.727 & 0.0241 & 22.0 & 15\\ [1ex]  \hline \\ [-1ex] 
JPBL & $0.284^{+0.008}_{-0.007}$ & $-5.1^{+1.7}_{-2.8}$ & $-1.8^{+0.9}_{-1.4}$ & $-0.63^{+0.45}_{-0.28}$ & 0.719 & 0.0241 & 720.7 & 755\\ [1ex]  \hline \\ [-1ex] 
All  & $0.275^{+0.006}_{-0.006}$ & $-4.1^{+0.5}_{-0.9}$ & $-1.5^{+0.2}_{-0.4}$ & $-0.78^{+0.13}_{-0.06}$ & $0.736$ & $0.0240$ & 731.9 & 767 \\ [1ex] 
  \hline \hline \\ [-1ex] 
All \citetalias{bib:neveu14} & $0.276^{+0.014}_{-0.009}$ & $-4.3^{+0.5}_{-1.1}$ & $-1.6^{+0.2}_{-0.6}$ & $-0.77^{+0.10}_{-0.06}$ & 0.726 & 0.0219 & 731.6 & 760\\ [1ex] \hline  
\end{tabular}
\tablefoot{The JLA SN~Ia sample is used with systematics included; $\alpha$ and $\beta$ are fixed to their marginalised values. $h$ and $\Omega_b^0h^2$ have been minimised so no uncertainties are provided. The best fit $\chi^2$ for the \citetalias{bib:neveu14} analysis was reevaluated to account for an error in the SN~Ia $\chi^2$ contribution.} 
\end{center}
\end{table*}

\begin{figure*}[hbtp]
\begin{center}
\epsfig{figure=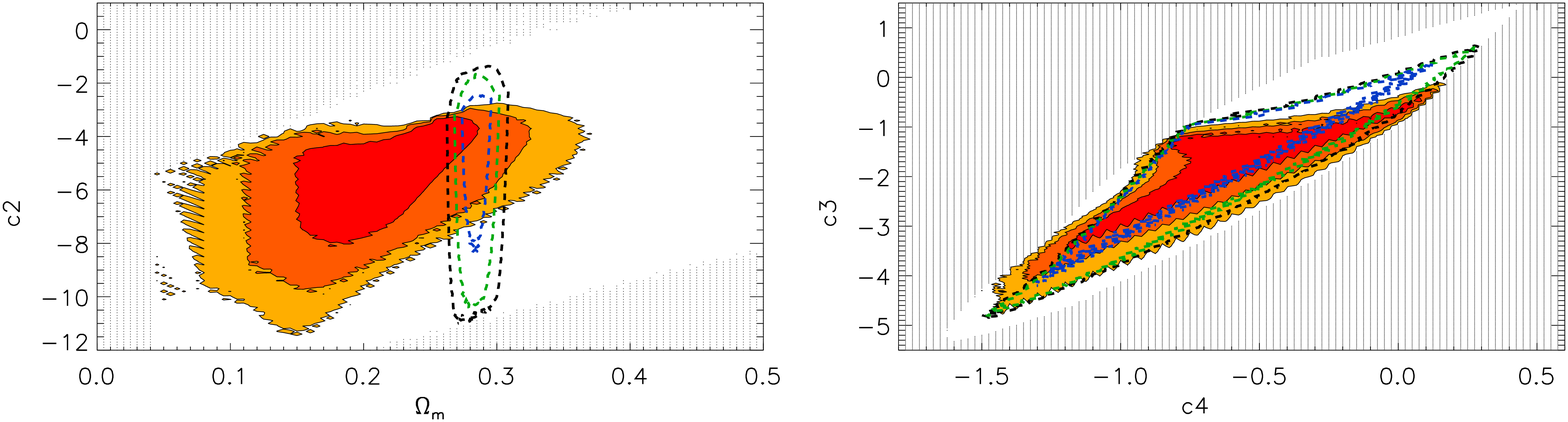, width=\textwidth} 
\epsfig{figure=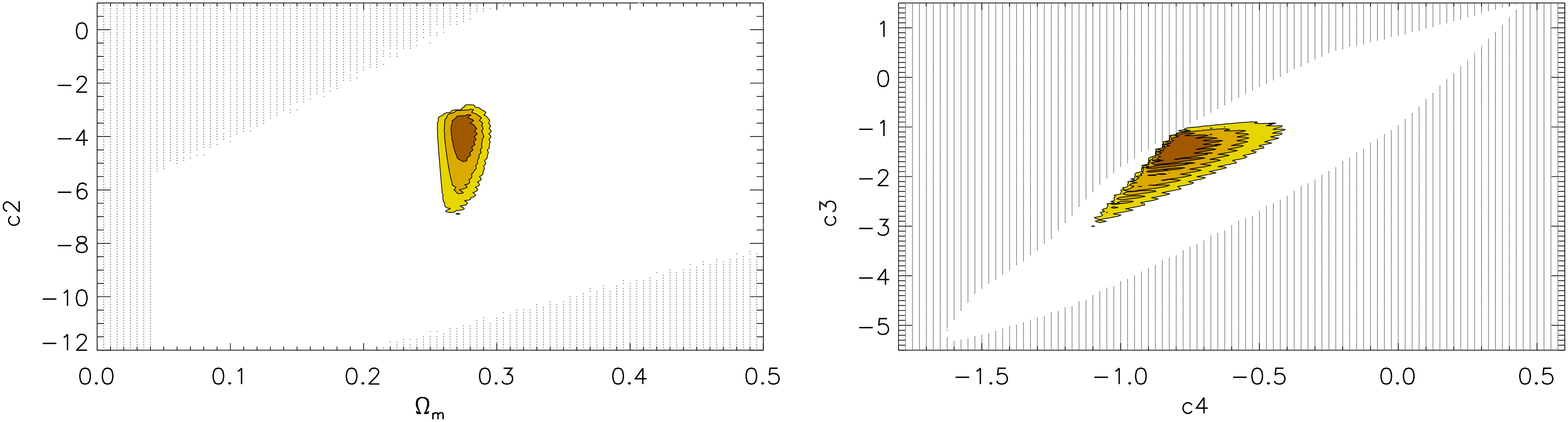, width=\textwidth} 
\caption[]{Top: Cosmological constraints on the uncoupled Galileon model from growth data (red) and from JLA+\textit{Planck}+BAO+Ly$\alpha$ data (dashed). Bottom: Combined constraints on the uncoupled Galileon model from all data combined. The filled dark, medium and light coloured contours enclose 68.3, 95.4 and 99.7\% of the probability, respectively. Dark dotted regions correspond to scenarios rejected by theoretical constraints. Only two projections out of six are shown. Results in the other projections are similar.} 
\label{fig:gof_jlabaoplancklyalpha}
\end{center}
\end{figure*}

\subsubsection{Scalar perturbations}

\par As suggested by \cite{bib:appleby}, outside the quasi-static approximation the propagation equation for $\delta y$ leads to two conditions, which we again checked from $z=0$ to $z=10^7$ to ensure the viability of the linearly perturbed model:
\begin{enumerate}
\item a no-ghost condition, which requires a positive energy for the perturbation
\begin{equation}\label{eq:noghost1}
\kappa_2 + \frac{3}{2} \frac{\kappa_5^2}{\kappa_4}<0 ;
\end{equation}
\item a Laplace stability condition for the propagation speed of the perturbed field
\begin{equation}\label{eq:cs1}
c_s^2=\frac{4\kappa_1\kappa_4\kappa_5 - 2\kappa_3 \kappa_5^2 - 2 \kappa_4^2 \kappa_6}{\kappa_4(2\kappa_4\kappa_2 + 3\kappa_5^2)}>0
\end{equation}
\end{enumerate}
with
\begin{equation}
\kappa_2 = -\frac{\bar c_2}{2}+6\bar c_3 \h^3\x -27 \bar c_4 \h^4 \x^2 +30 \bar c_5 \h^6 \x^3 +2\bar c_G \h^2.
\end{equation}

\begin{table*}[htb]
\caption[]{Disformally coupled Galileon model best fit values from JPBL data alone and combined with growth rate measurements.}
\label{tab:resultscG}
\begin{center}
\begin{tabular}{cccccccccc} \hline \hline \\ [-1ex]
Probe & $\Omega_m^0$ & $\bar c_2$ & $\bar c_3$ & $\bar c_4$ & $\bar c_G$ & $h$ & $\Omega_b^0h^2$ & $ \chi^2$  & $N_{\rm data}$  \\  [1ex] \hline \\ [-1ex]
JPBL & $0.288^{+0.009}_{-0.007}$ & $-3.1^{+1.8}_{-1.6}$ & $0.1^{+1.6}_{-1.2}$ & $0.20^{+0.65}_{-0.63}$ & $0.69^{+0.55}_{-0.46}$ & 0.710 & 0.0244 & 721.1 & 755 \\ [1ex]  \hline\\ [-1ex] 
All & $0.280^{+0.007}_{-0.005}$ & $-3.4^{+0.4}_{-0.7}$ & $-1.1^{+0.2}_{-0.3}$ & $-0.61^{+0.09}_{-0.09}$ & $0.15^{+0.09}_{-0.06}$ & 0.727 & 0.0240 & 724.7 & 767 \\ [1ex]  \hline \hline \\ [-1ex] 
All \citetalias{bib:neveu14} & $0.279^{+0.013}_{-0.008}$ & $-3.4^{+0.3}_{-0.6}$ & $-1.0^{+0.2}_{-0.3}$ & $-0.61^{+0.09}_{-0.08}$ & $0.15^{+0.08}_{-0.06}$ & 0.719 & 0.0220 & 722.8 & 760\\ [1ex]  \hline
\end{tabular}
\tablefoot{The best fit $\chi^2$ for the \citetalias{bib:neveu14} analysis was reevaluated to account for an error in the SN~Ia $\chi^2$ contribution.}
\end{center}
\end{table*}

\begin{figure*}[hbtp]
\begin{center}
\epsfig{figure=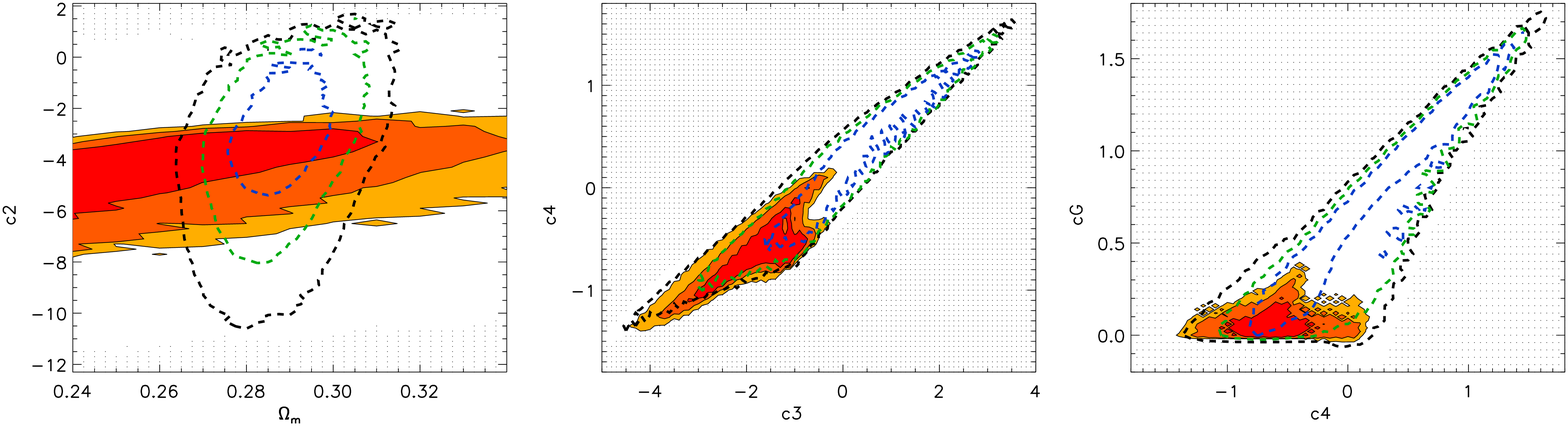, width=\textwidth} 
\epsfig{figure=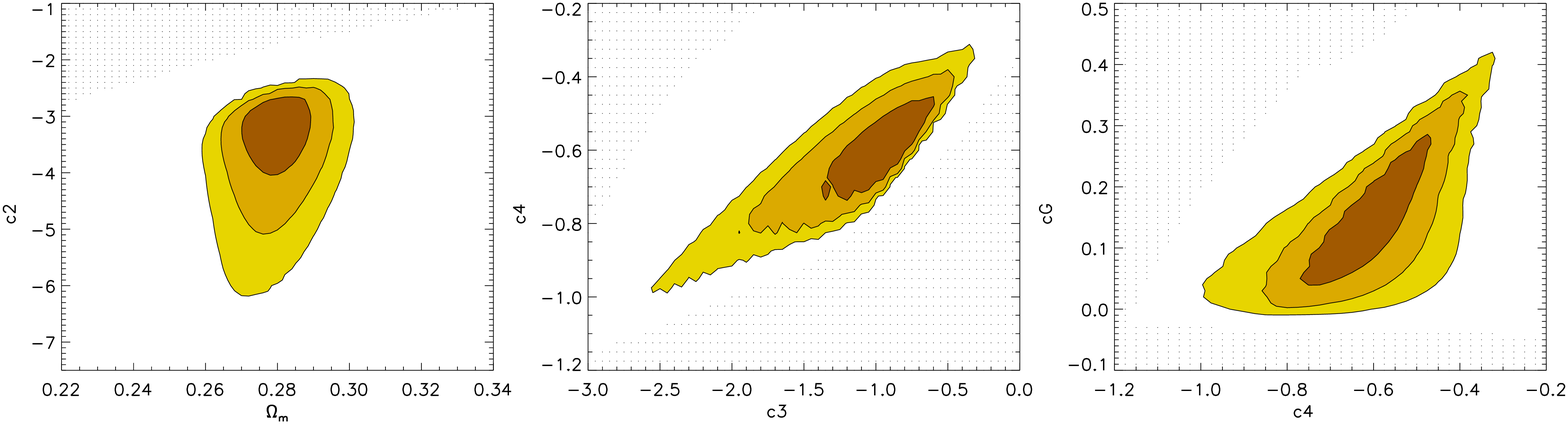, width=\textwidth} 
\caption[]{Top: Cosmological constraints on the disformally coupled Galileon model from growth data (red) and from JLA+\textit{Planck}+BAO+Ly$\alpha$ data (dashed). Bottom: Combined constraints on the disformally coupled Galileon model from all data combined. The filled dark, medium and light coloured contours enclose 68.3, 95.4 and 99.7\% of the probability, respectively. Dark dotted regions correspond to scenarios rejected by theoretical constraints. Three projections out of ten are shown.} 
\label{fig:gof_all_combined_cg}
\end{center}
\end{figure*}

\subsubsection{Tensorial perturbations}\label{sec:tensorial}

We also added two conditions derived by \citet{bib:felice2011} for the propagation of tensor perturbations. Considering a traceless and divergence-free perturbation $\delta g_{ij} = a^2h_{ij}$, these authors obtained identical perturbed actions at second order for each of the two polarisation modes $h_\oplus$ and $h_\otimes$.  For $h_\oplus$,
\begin{equation}
\delta S_T^{(2)}=\frac{1}{2}\int dtd^3xa^3Q_T\left[\dot h^2_\oplus - \frac{c_T^2}{a^2}(\nabla h_\oplus)^2 \right]
\end{equation}
with $Q_T$ and $c_T$ as defined below. From that equation, we extracted two conditions in our parametrisation that have to be satisfied (again from $z=0$ to $z=10^7$):
\begin{enumerate}
\item a no-ghost condition
\begin{equation}\label{eq:noghost2}
\frac{Q_T}{M_P^2} = \frac{1}{2} - \frac{3}{4}\bar c_4 \h^4\x^4 + \frac{3}{2}\bar c_5 \h^5 \x^5 + \frac{\bar c_G}{2}\h^2\x^2 - \bar c_0 \y > 0;
\end{equation}
\item a Laplace stability condition
\begin{equation}\label{eq:cs2}
c_T^2=\frac{\frac{1}{2} +  \frac{\bar c_4}{4} \h^4\x^4 + \frac{3}{2}\bar c_5 \h^5 \x^4(\h\x)'- \frac{\bar c_G}{2}\h^2\x^2 - \bar c_0 \y}{\frac{1}{2} - \frac{3}{4}\bar c_4 \h^4\x^4 + \frac{3}{2}\bar c_5 \h^6 \x^5+ \frac{\bar c_G}{2}\h^2\x^2 - \bar c_0 \y}>0.
\end{equation}
\end{enumerate}

All these conditions allowed us to reduce the Galileon parameter space significantly.

\begin{figure*}[htb]
\begin{center}
\epsfig{figure=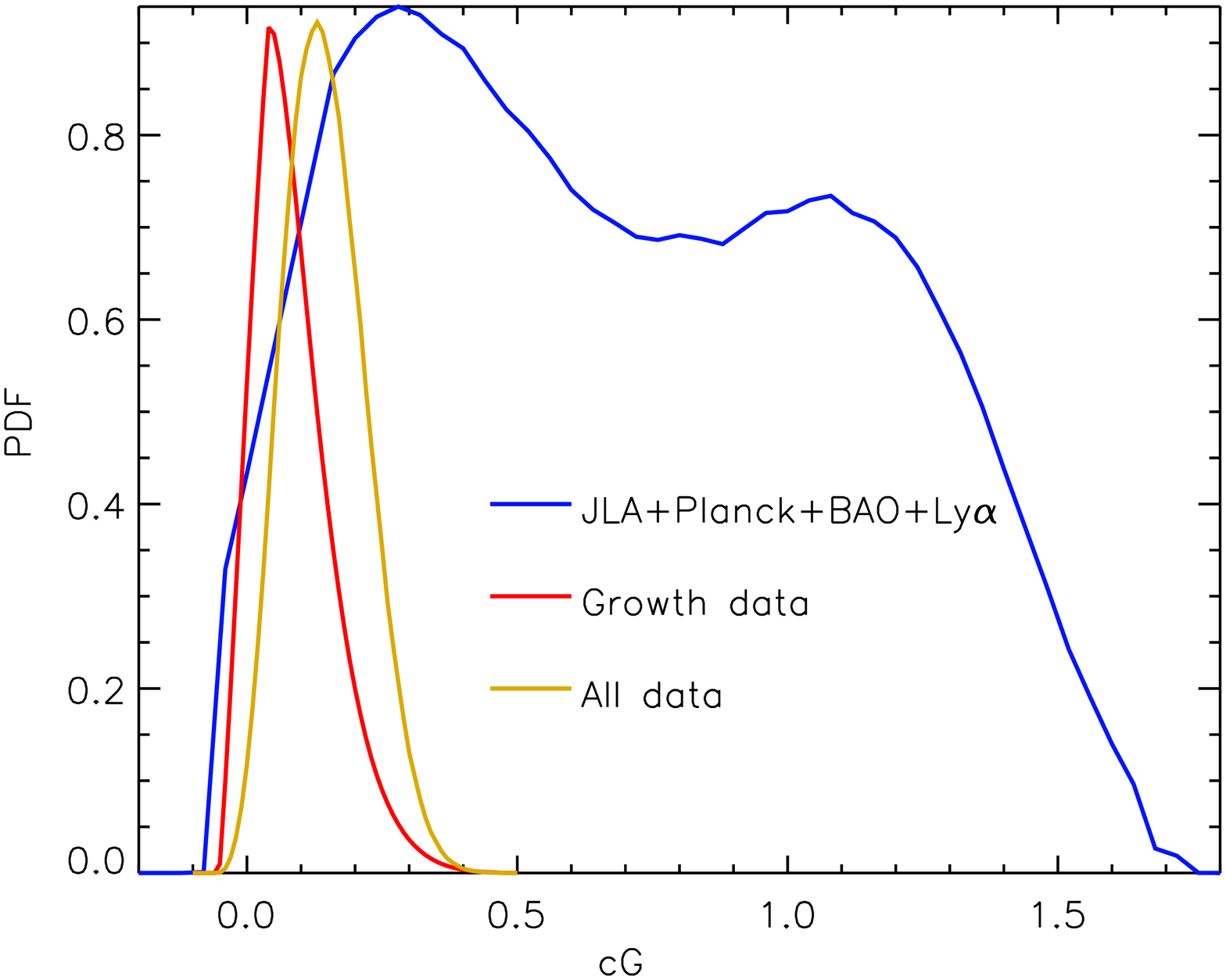, width=\columnwidth} 
\epsfig{figure=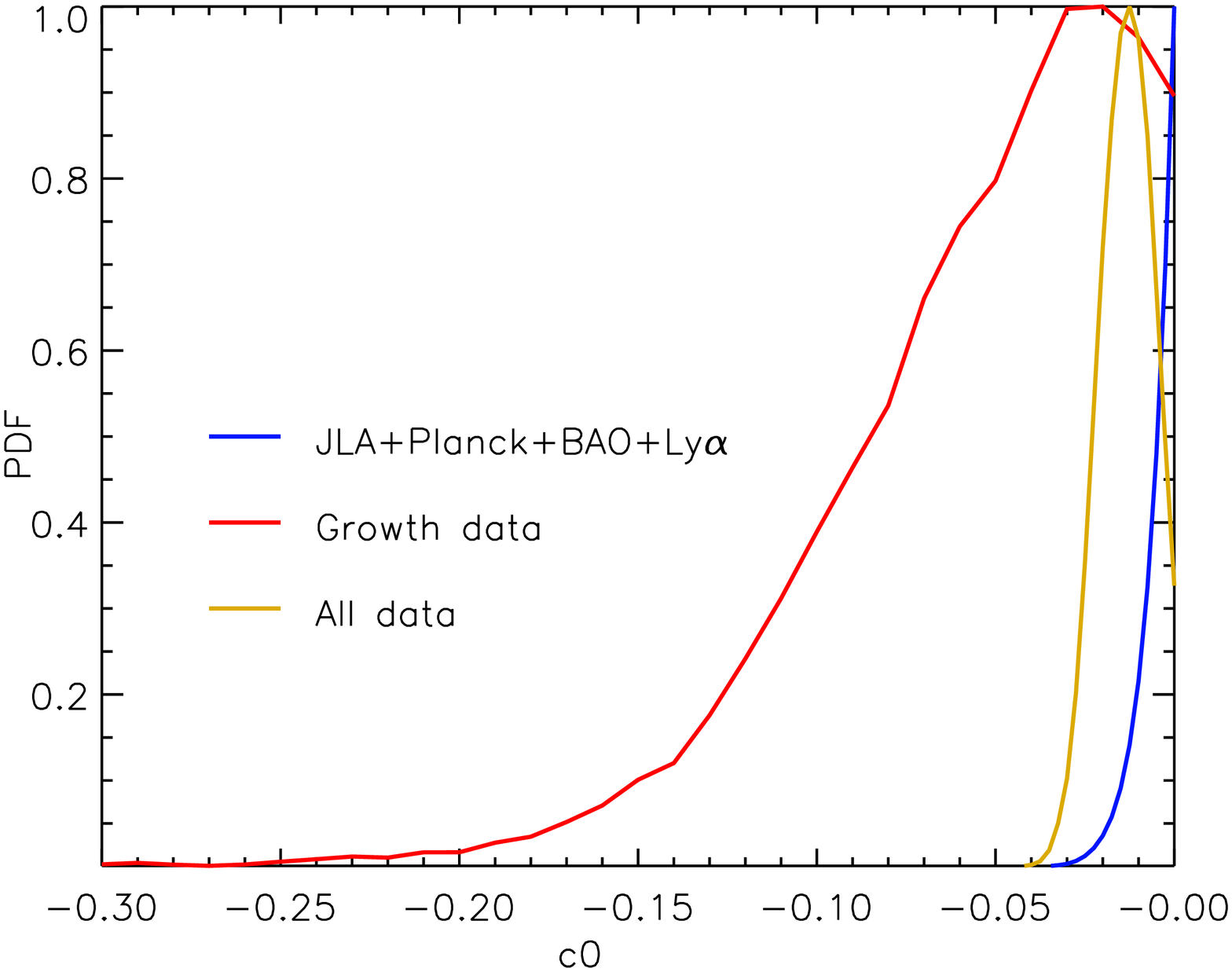, width=\columnwidth} 
\caption[]{Left: Disformally coupled Galileon mode $\bar c_G$ probability density functions obtained with different data sets, marginalising over all other parameters. Right: Conformally coupled Galileon mode $\bar c_G$ probability density functions obtained with different data sets, marginalising over all other parameters. For clarity, probability density function maxima are normalised to one.} 
\label{fig:gof_all_combined_pdf}
\end{center}
\end{figure*}

\begin{figure*}[hbtp]
\begin{center}
\epsfig{figure=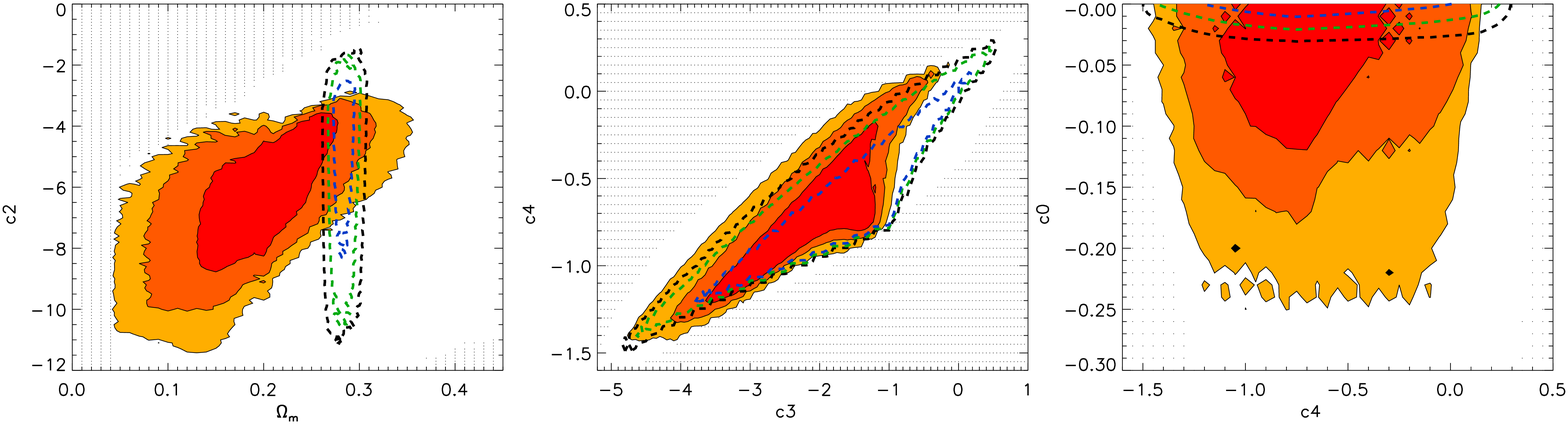, width=\textwidth} 
\epsfig{figure=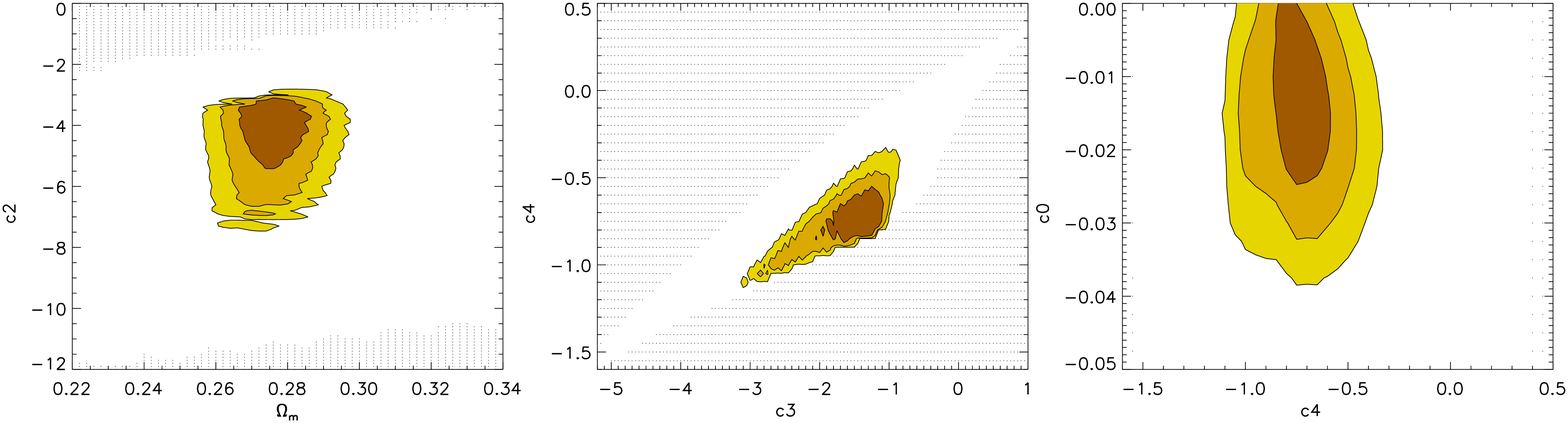, width=\textwidth} 
\caption[]{Top: Cosmological constraints on the conformally coupled Galileon model from growth data (red) and from JLA+\textit{Planck}+BAO+Ly$\alpha$ data (dashed). Bottom: Combined constraints on the conformally coupled Galileon model from all data combined. The filled dark, medium and light yellow contours enclose 68.3, 95.4 and 99.7\% of the probability, respectively. Dark dotted regions correspond to scenarios rejected by theoretical constraints. Only three projections out of 10 are shown.}
\label{fig:gof_all_combined_c0}
\end{center}
\end{figure*}

\begin{table*}[htb]
\caption[]{Conformally coupled Galileon model best fit values from JPBL data alone and combined with growth rate measurements.}
\label{tab:resultsc0}
\begin{center}
\begin{tabular}{cccccccccc} \hline \hline \\ [-1ex]
Probe & $\Omega_m^0$ & $\bar c_2$ & $\bar c_3$ & $\bar c_4$ & $\bar c_0<0$ & $h$ & $\Omega_b^0h^2$ & $ \chi^2$  & $N_{\rm data}$ \\  [1ex] \hline \\ [-1ex]
JPBL  & $0.284^{+0.008}_{-0.006}$ & $-5.1^{+1.7}_{-2.8}$ & $-1.8^{+0.9}_{-1.4}$ & $-0.63^{+0.47}_{-0.28}$ & $-0.017$ ($95\%$ CL) & 0.719 & 0.0241 & 720.2 & 755 \\ [1ex]  \hline\\ [-1ex] 
All & $0.276^{+0.007}_{-0.005}$ & $-4.4^{+0.6}_{-1.4}$ & $-1.6^{+0.3}_{-0.7}$ & $-0.74^{+0.16}_{-0.08}$ & $-0.013^{+0.008}_{-0.008}$ & 0.747 & 0.0244 & 730.6 & 767\\ [1ex]  \hline
\end{tabular}
\tablefoot{With JPBL data, the $\bar c_0$ probability density function is maximum at $0$. As $\bar c_0 > 0$ values are forbidden, only a lower bound is quoted in this line, and the best fit values of the other parameters are evaluated assuming $\bar c_0=0$.}
\end{center}
\end{table*}

\begin{figure*}[hbtp]
\begin{center}
\epsfig{figure=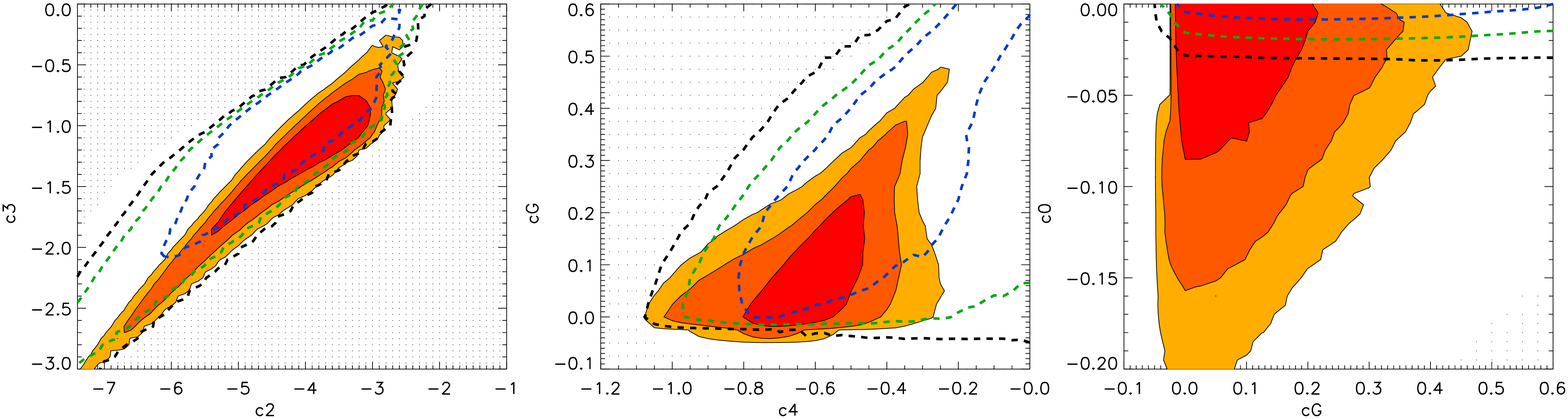, width=\textwidth} 
\epsfig{figure=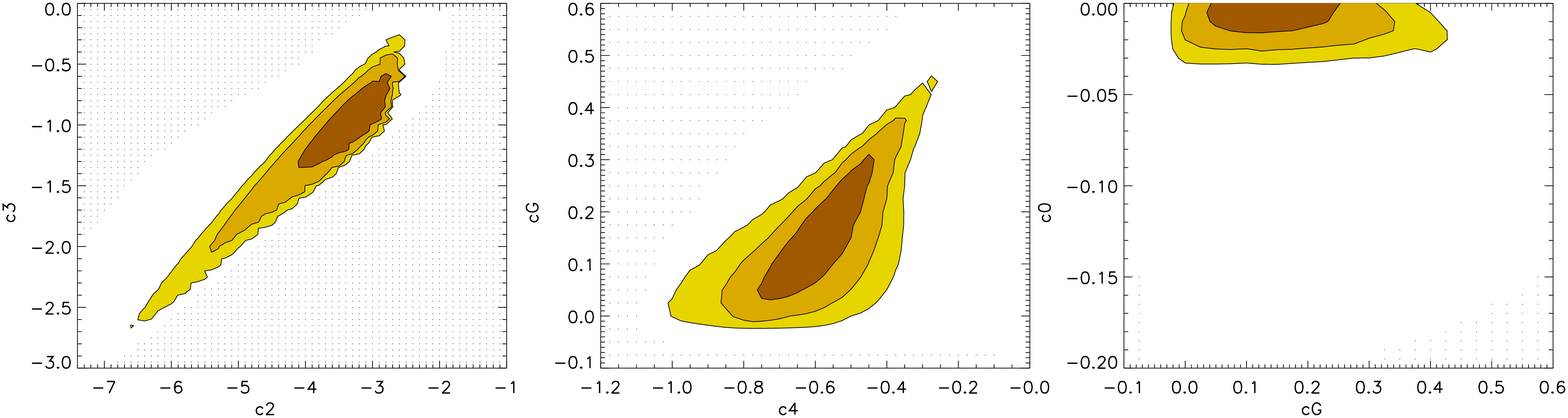, width=\textwidth} 
\caption[]{Top: Cosmological constraints on the Galileon model coupled conformally and disformally to matter from growth data (red) and from JLA+\textit{Planck}+BAO+Ly$\alpha$ data (dashed). Bottom: Combined constraints on the Galileon model coupled conformally and disformally to matter from all data combined. The filled dark, medium and light yellow contours enclose 68.3, 95.4 and 99.7\% of the probability, respectively. Dark dotted regions correspond to scenarios rejected by theoretical constraints. In all results, $\Omega_m^0$ was fixed to 0.28. Only three projections out of ten are shown.}
\label{fig:gof_all_combined_c0cG}
\end{center}
\end{figure*}

\begin{table*}[htb]
\caption[]{Galileon model conformally and disformally coupled to matter best fit values from JPBL data alone and combined with growth rate measurements, with fixed matter density $\Omega_m^0=0.28$.}
\label{tab:resultsc0cG}
\begin{center}
\begin{tabular}{cccccccccc} \hline \hline \\ [-1ex]
Probe & $\bar c_2$ & $\bar c_3$ & $\bar c_4$ & $\bar c_G$ & $\bar c_0$ & $h$ & $\Omega_b^0h^2$ & $ \chi^2$ & $N_{\rm data}$  \\  [1ex] \hline \\ [-1ex]
All & $-3.4^{+0.4}_{-0.7}$ & $-1.0^{+0.2}_{-0.3}$ & $-0.61^{+0.10}_{-0.06}$ & $0.15^{+0.10}_{-0.06}$ & $-0.027$ ($95\%$ CL) & 0.727 & 0.0240 & 724.6 & 767 \\ [1ex]  \hline
\end{tabular}
\tablefoot{The $\bar c_0$ probability density function is maximum at $0$. As $\bar c_0 > 0$ values are forbidden, only a lower bound is quoted, and the best fit values of the other parameters are evaluated assuming $\bar c_0=0$.}
\end{center}
\end{table*}

\section{Results}\label{sec:results}

In this section, we present new cosmological constraints on the $\bar c_i$ parameters of the Galileon model with different couplings to matter. Results are   discussed further in Section~\ref{sec:disc}. 

To appreciate the goodness of fit of the Galileon models, it is better to avoid  reduced $\chi^2$ values because they require the number of degrees of freedom $N_{\rm dof}$ to be determined. As explained in \cite{bib:andrae}, the usual definition of $N_{\rm dof}$ as the difference between the number of measurements and the number of free independent parameters, is generally not valid in the case of non-linear models, nor 
when the parameter space is limited by priors like the theoretical conditions we presented in section~\ref{sec:theoconstraints}. In our papers, we thus prefer to compare values of the best fit $\chi^2$ to the number of measurements $N_{\rm data}$, as the total $\chi^2$ gives an estimate of the mean size of the residuals between predictions and data. This technical point is  discussed further in Appendix~\ref{sec:goodness}. A summary of the best fit $\chi^2$ and  $N_{\rm data}$ values for each cosmological probe  and all tested models is given in Section~\ref{sec:disc}.


\subsection{Uncoupled Galileon model}\label{sec:uncoupled}

New constraints on the uncoupled Galileon model are presented in Table~\ref{tab:results} and Figure~\ref{fig:gof_jlabaoplancklyalpha}. Results are very similar to those obtained with previous data sets in \citetalias{bib:neveu} and \citetalias{bib:neveu14} as shown in the last row of Table~\ref{tab:results}. Although there is  more data, the final $\chi^2$ is nearly unchanged because there is  better agreement with data, especially growth rate measurements. We note   that the use of \Planck priors containing polarisation information brought tighter constraints on $\Omega_m^0$ as already observed with the standard cosmological models (see Section~\ref{sec:standard}).

\begin{figure*}[hbtp]
\begin{center}
\epsfig{figure=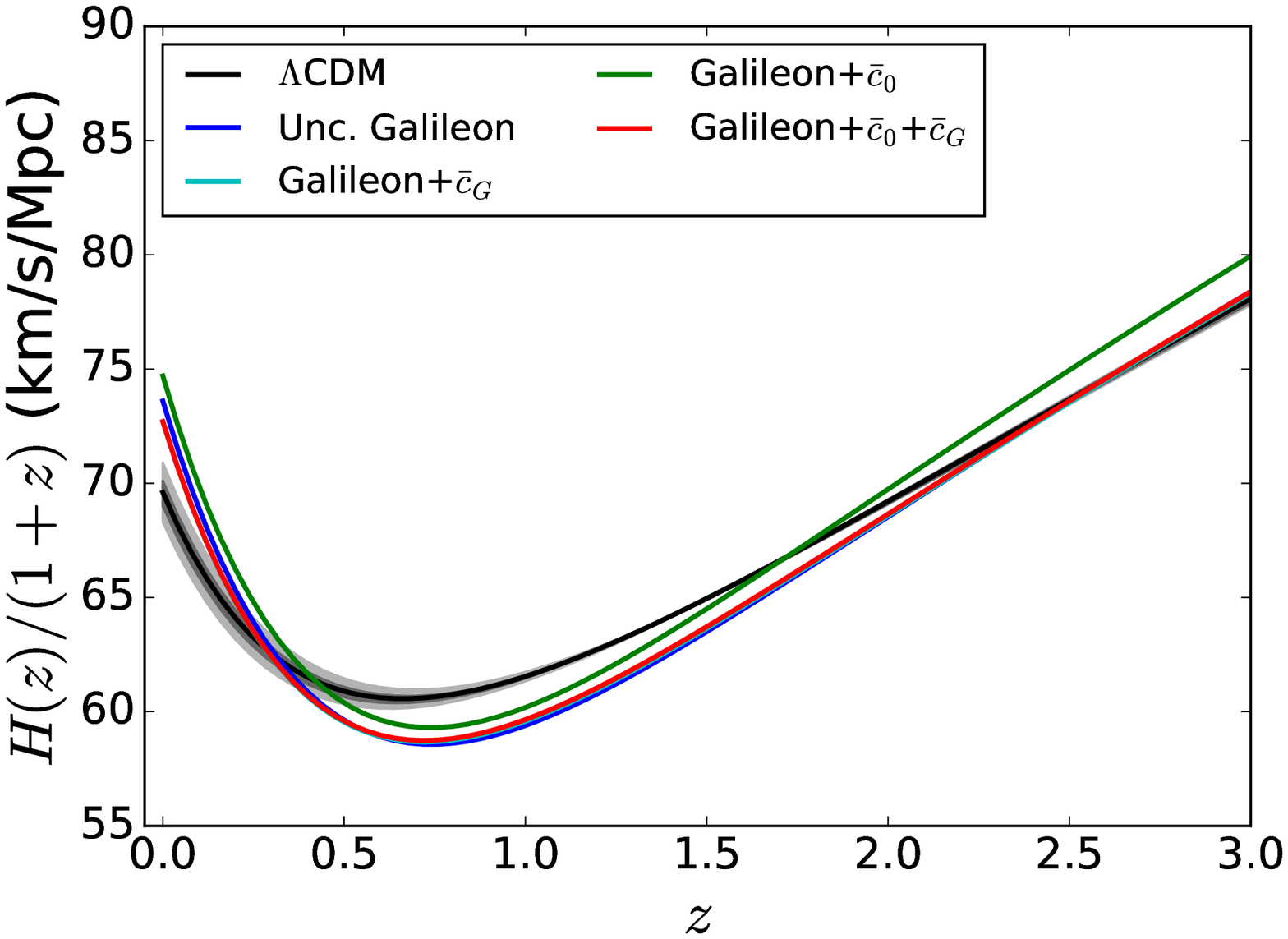, width=0.67\columnwidth} 
\epsfig{figure=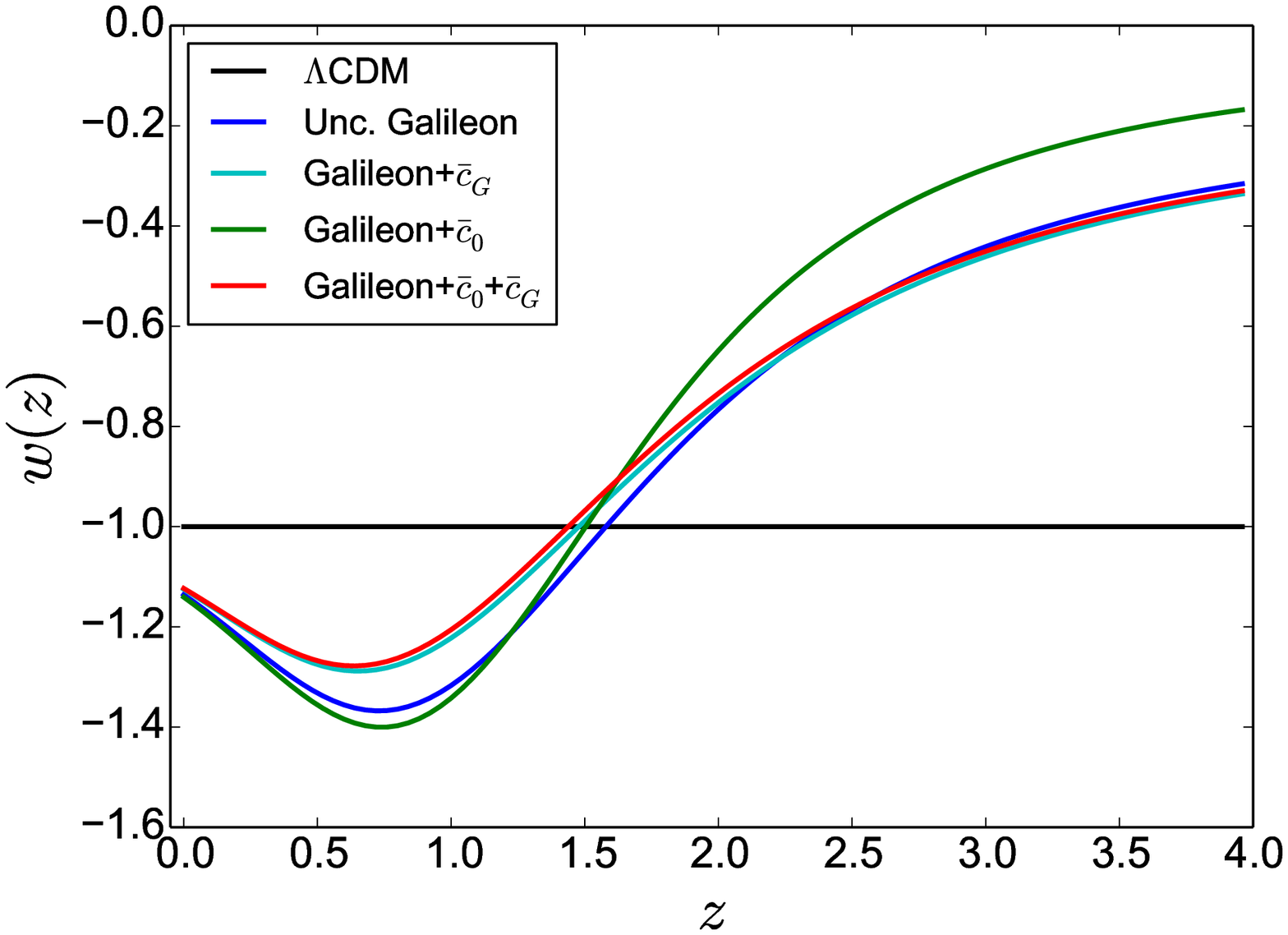, width=0.67\columnwidth} 
\epsfig{figure=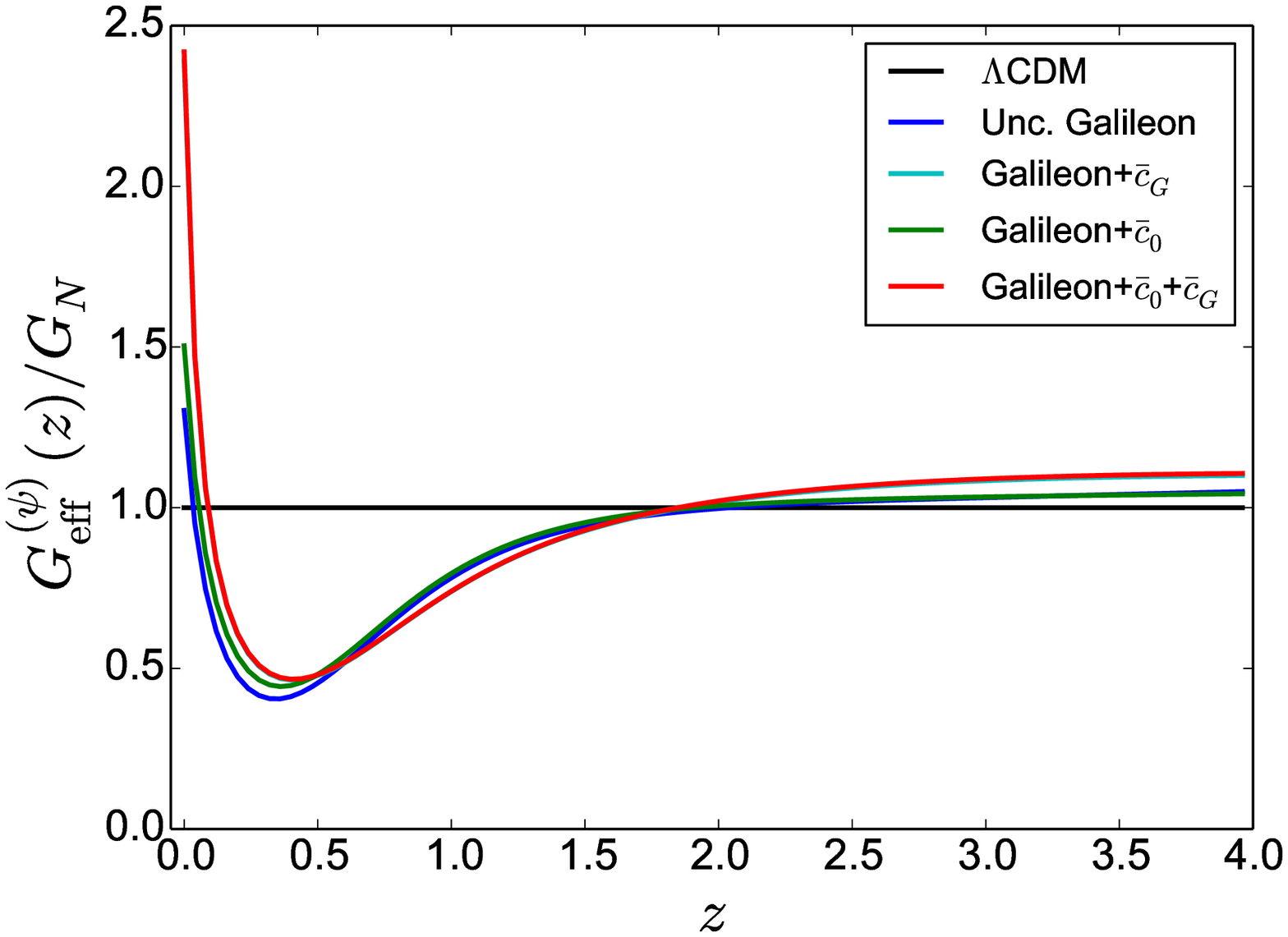, width=0.67\columnwidth} 
\caption[]{Expansion speed parameter $H(z)/(1+z)$ (left), effective equation of state parameter $w(z)$ (middle) and gravitational coupling $G_{\mathrm{eff}}^{(\psi)}(z)$ (right) as a function of redshift in the best fit scenarios of the four Galileon models studied in this paper, compared with $\Lambda$CDM best fit predictions (with 68\% and 95\% confidence ranges indicated as grey bands, whenever relevant).} 
\label{fig:w}
\end{center}
\end{figure*}

\subsection{Galileon model disformally coupled to matter}\label{sec:coupled}

New constraints on the disformally coupled Galileon model are presented in Table~\ref{tab:resultscG} and Figure~\ref{fig:gof_all_combined_cg}. Results are very similar to those obtained with previous data sets in \citetalias{bib:neveu14} as shown in the last row of Table~\ref{tab:resultscG}. The final $\chi^2$ is still better than in the uncoupled case. Moreover, the $\bar c_G$ constraint, $\bar c_G = 0.15^{+0.09}_{-0.06}$, excludes a null coupling at the $2.3\sigma$ level. With the new growth rate measurements, the tension between growth data and distance measurements decreased and the agreement between the Galileon best fit scenario and growth data alone improved.

Figure~\ref{fig:gof_all_combined_cg} shows that for the $\bar c_i$s contours from distance measurements are  only determined by the theoretical conditions. The probability density functions for these  parameters are wide as illustrated in Figure~\ref{fig:gof_all_combined_pdf}. Only growth data brought constraints on the $\bar c_i$ parameters, in particular on $\bar c_G$. 

The reason why distance measurements lead to flat probability density functions over the available $\bar c_i$ parameter space may have a more fundamental origin.  Indeed, the Galileon theory belongs to the Horndeski model class and it has been shown that these models are invariant under disformal transformations \citep{bib:bettoni}. Moreover, \cite{bib:zumalacarregui} has shown that a disformally coupled theory in which the gravitational sector has the Einstein-Hilbert form is equivalent to a quartic Dirac-Born-Infeld Galileon Lagrangian \cite{bib:deRhamDBI}. Thus, a disformally coupled Galileon model belongs to a wide class of equivalent disformal models which may also present  equivalent expansion histories that cosmological distances are not able to distinguish. The wide contours in Figure~\ref{fig:gof_all_combined_cg} in the allowed $\bar c_i$ parameter space may be related to this theoretical property.

\subsection{Galileon model conformally coupled to matter}\label{sec:conform}

For the first time, we explored the parameter space of the Galileon model conformally coupled to matter. First, we found that the region $\bar c_0 > 0$ is forbidden by theoretical constraints, in particular by equation~\ref{eq:cs1}. In the following, only $\bar c_0<0$ is explored. The cosmological constraints are presented in Table~\ref{tab:resultsc0} and Figure~\ref{fig:gof_all_combined_c0}. 

Contrary to the disformal coupling, the conformal coupling is tightly constrained by the JPBL data set. Growth data are less severe and allow large  non-zero values for this coupling to matter. As the maximum of the $\bar c_0$ probability density function is zero  with the JPBL data set, only a 95\% confidence level limit is set on this parameter. However, when combining with growth data, a non-zero value is preferred at the $1.6\sigma$ level only, as also shown in Figure~\ref{fig:gof_all_combined_pdf}. The level of agreement between data and predictions is similar to that in the uncoupled case, even though    one parameter has been added. The conformal coupling does not appear to be favoured by the data, at least at the present level of  precision.

\subsection{Galileon model conformally and disformally coupled to matter}\label{sec:conformdis}

Finally, we also tested a Galileon model with both disformal and conformal couplings to matter. The parameter space is so wide that, for computational reasons, we imposed a fixed value $\Omega_m^0 = 0.28$, in agreement with all previous Galileon constraints. Results are presented in  Table~\ref{tab:resultsc0cG} and Figure~\ref{fig:gof_all_combined_c0cG}. 

The conclusions are very similar to the previous ones. A null conformal coupling and a non-zero disformal coupling are preferred. We observed no new interplay between the additional coupling terms in the theory leading to differences in the best fit scenarios. The best fit and $\chi^2$ values are completely equivalent to those of the disformally coupled Galileon model. 

\section{Discussion}\label{sec:disc}

\subsection{Comparing the Galileon models}

\begin{figure}[hbtp]
\begin{center}
\epsfig{figure=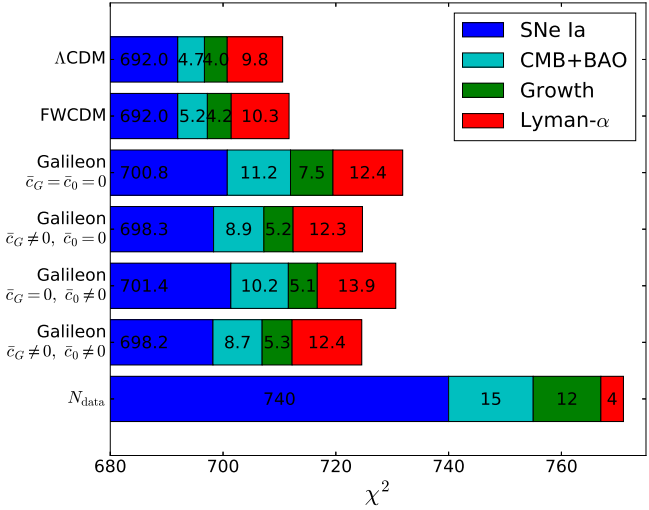, width=\columnwidth} 
\caption{$\chi^2$ values for the best fit cosmological models constrained with all data. The breakdown of the total $\chi^2$ between the different data sets is indicated as JLA (blue), BAO+\Planck (cyan), growth data (green), and Lyman-$\alpha$ (red). The  $N_{\rm data}$  line shows the number of measurements from each data set.} 
\label{fig:chi2}
\end{center}
\end{figure}

The particular behaviour of the Galileon best fit scenarios is displayed in Figure~\ref{fig:w}. For the four cases, the equation of state parameter $w(z)$ deviates strongly from $w_\Lambda = -1$ between $z=0$ and $z=3$ before converging toward a value close to $0$ in the early universe as already shown in \citetalias{bib:neveu} and \citetalias{bib:neveu14}. The effective gravitational coupling $G_{\mathrm{eff}}^{(\psi)}(z)$ also deviates from $G_N$ in the late universe. We note   that $G_{\mathrm{eff}}^{(\psi)}(z)$ determines the gravitational strength that rules structure formation at large scales and there is no reason why this coupling should be $G_N$ as measured in local experiments. The introduction of the Galileon couplings to matter leads to best fit scenarios which are rather similar to the uncoupled best fit scenario. 

A summary of the different $\chi^2$ values obtained with the full data set is given in Figure~\ref{fig:chi2}. In each case, the contribution from the different probes is detailed. It confirms that a disformal coupling  improves the agreement between data and the Galileon model,
whether the Galileon field is also conformally coupled or not. 
This histogram also shows that if   the individual $\chi^2$ values are compared with the number of measurements $N_{\rm data}$, except the two Ly$\alpha$ measurements, no probe is particularly in tension with the Galileon model. But the Ly$\alpha$ probe exhibits similar tensions in the standard cosmological models as also revealed in many other works \citep{bib:delubac,bib:fontribera,bib:aubourg,bib:planck15}.

\subsection{Comparing the Galileon models with $\Lambda$CDM}

\subsubsection{Distance measurements}

\begin{figure}[h]
\begin{center}
\epsfig{figure=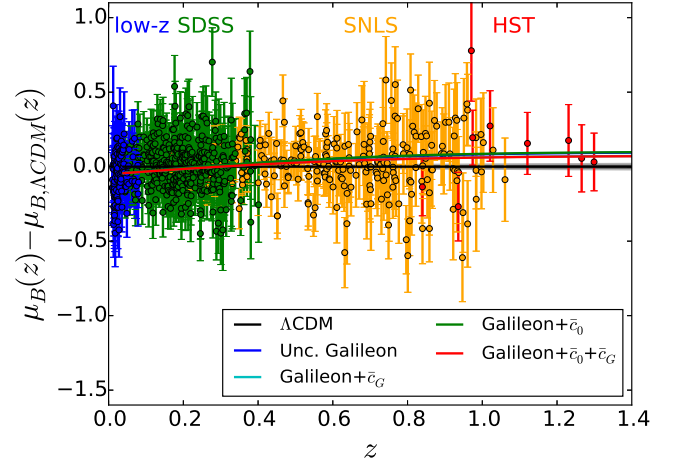, width=\columnwidth} 
\caption[]{Comparison of the best fit Galileon and $\Lambda$CDM SN~Ia magnitude predictions with data. The grey bands show the 68\% and 95\% confidence ranges allowed for the $\Lambda$CDM prediction using all data. The JLA data points are from \cite{bib:jla}.}
\label{fig:hubble}
\end{center}
\end{figure}

\begin{figure}[h]
\begin{center}
\epsfig{figure=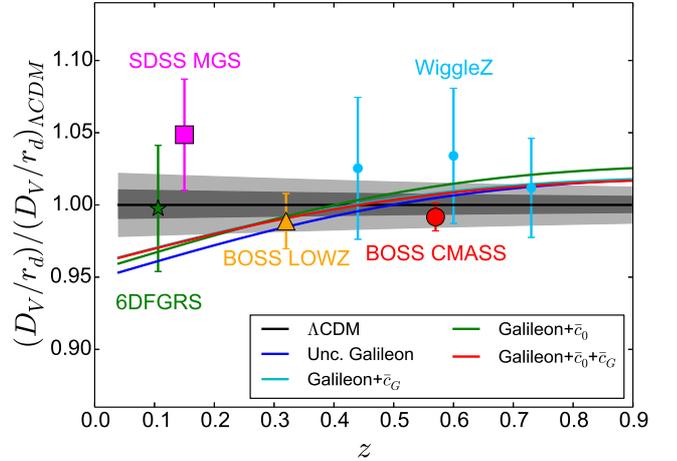, width=\columnwidth} 
\caption[]{Comparison of the best fit Galileon and $\Lambda$CDM $D_V / r_d$ predictions with data. The grey bands show the 68\% and 95\% confidence ranges allowed for the $\Lambda$CDM prediction using all data. The data points come from Table~\ref{tab:bao}. BOSS CMASS $D_V / r_d$ measurement from \cite{bib:boss14} is plotted for convenience, but is not used in the fitting procedure; the BOSS CMASS anistropic measurement is preferred. }
\label{fig:DV}
\end{center}
\end{figure}

\begin{figure*}[hbtp]
\begin{center}
\epsfig{figure=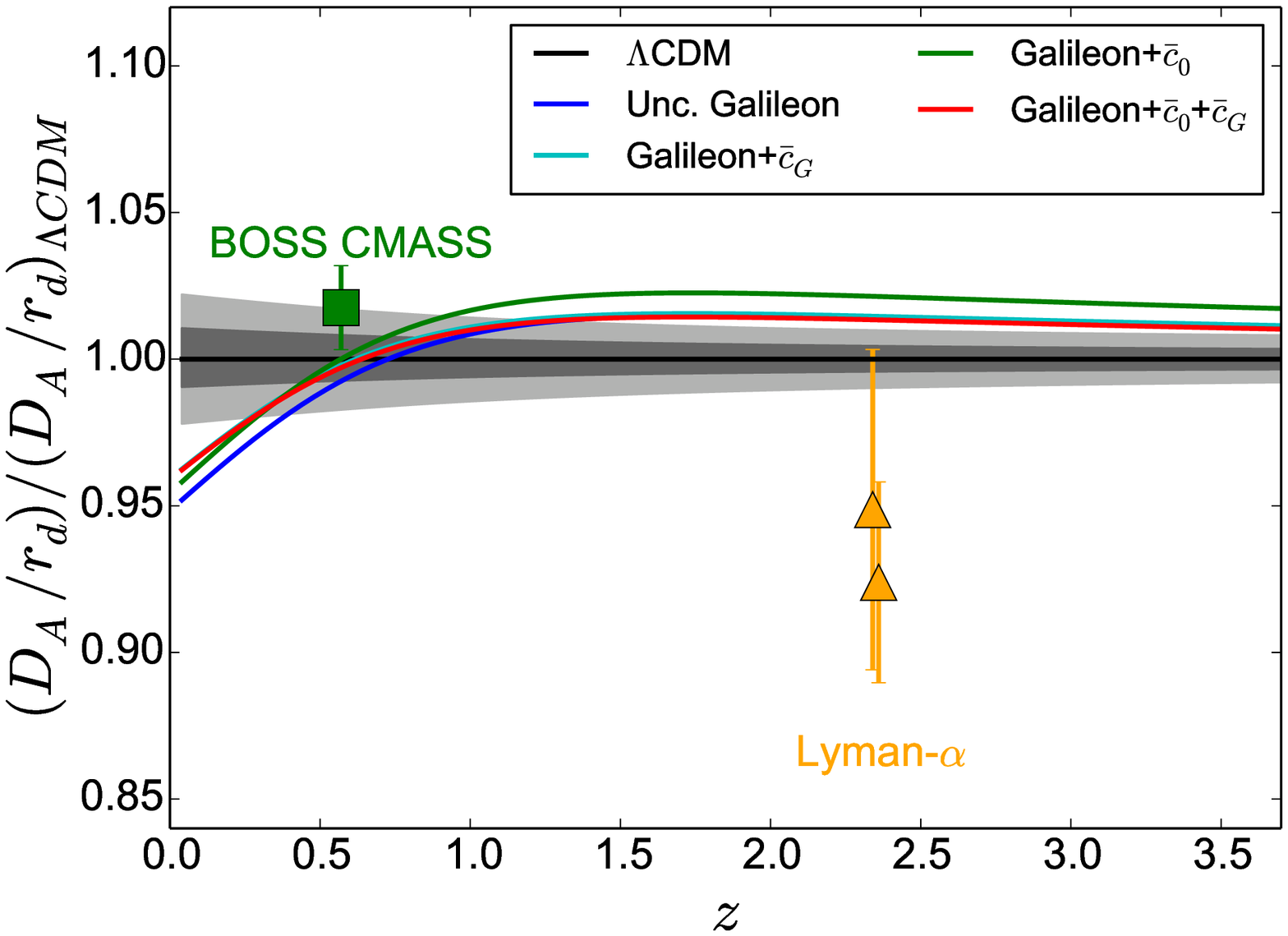, width=\columnwidth} 
\epsfig{figure=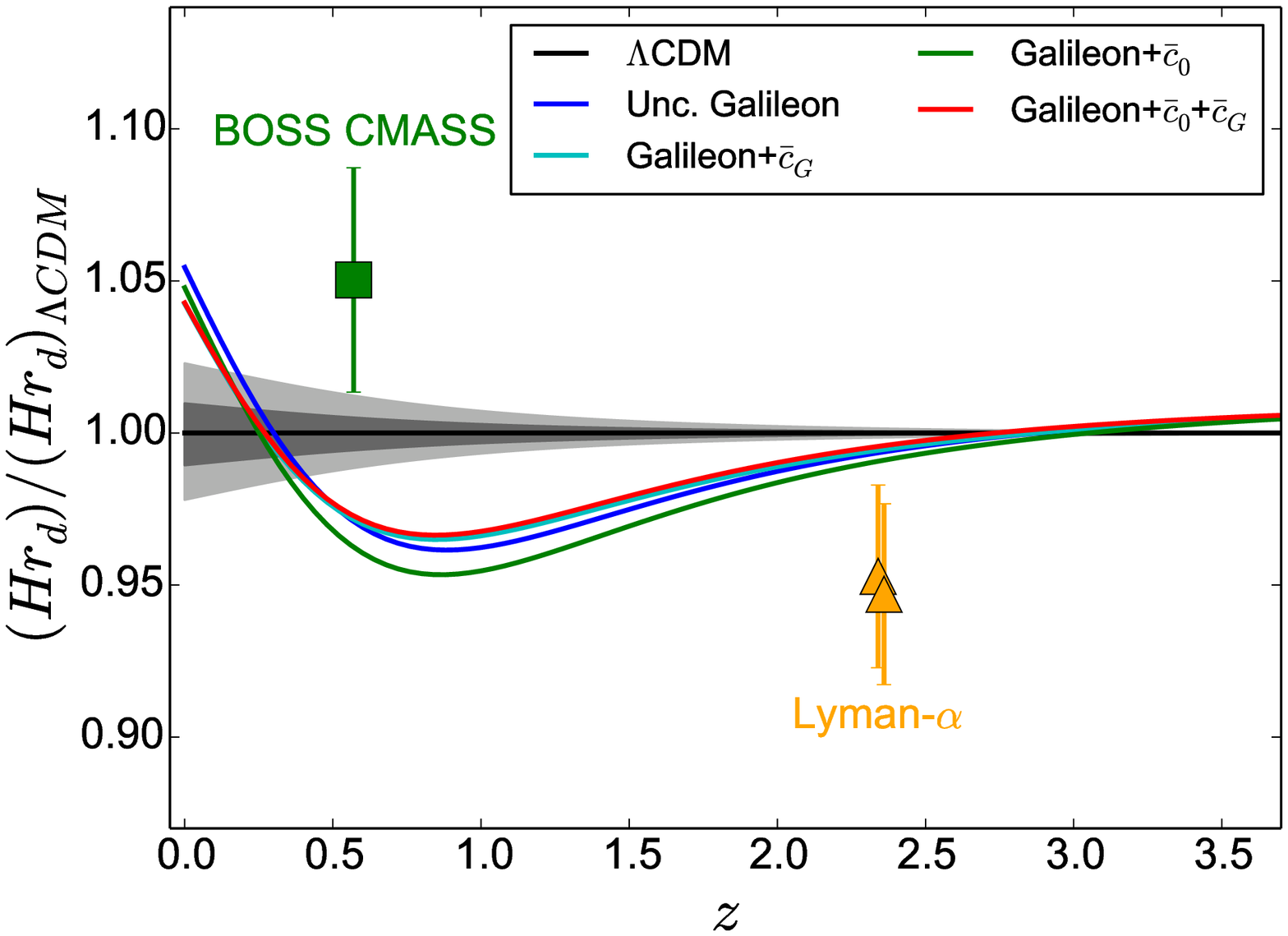, width=\columnwidth} 
\caption[]{Comparison of the best fit Galileon and $\Lambda$CDM BAO predictions with data. The grey bands show the 68\% and 95\% confidence ranges allowed for the $\Lambda$CDM prediction using all data. Left: $D_A / r_d$ predictions normalised to $\Lambda$CDM best fit scenario from Table~\ref{tab:lcdm} and compared to BOSS CMASS and Lyman-$\alpha$ measurements. Right: $H r_d$ predictions normalised to $\Lambda$CDM best fit scenario from Table~\ref{tab:lcdm} and compared to BOSS CMASS and Lyman-$\alpha$ measurements. } 
\label{fig:HDA}
\end{center}
\end{figure*}

The strong variations of the Galileon effective dark energy equation of state parameter $w(z)$ translates into a specific signature in the observables linked to cosmological distances. As shown in the left  plot in Figure~\ref{fig:w}, the deceleration of the expansion of the Universe and its re-acceleration after matter-dark energy transition at $z\approx 0.7$ are expected to be more important than in the $\Lambda$CDM model. 
The impact on the cosmological distance observables is shown in Figures~\ref{fig:hubble}, \ref{fig:DV}, and \ref{fig:HDA}.

The best fit SN~Ia magnitude predictions are plotted in Figure~\ref{fig:hubble} and are compared to JLA data. A deviation of about $0.08$\,mag is observed at redshift $z\approx 1$ between the $\Lambda$CDM and Galileon best fit scenarios, which corresponds to a deviation of 1.3\% on the luminosity distances. However, current SN~Ia measurements are limited by systematics, so larger supernova surveys will not help in discriminating between $\Lambda$CDM and the Galileon models unless calibration systematics improve.

\begin{figure}[!h]
\begin{center}
\epsfig{figure=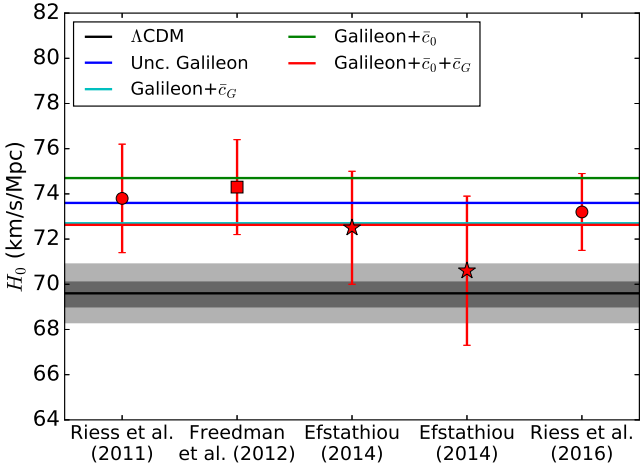, width=\columnwidth} 
\caption[]{Comparison of the best fit Galileon and $\Lambda$CDM $H_0$ predictions with direct measurements from Table~\ref{tab:h0}. These data points are not used in the cosmological fits. The grey bands show the 68\% and 95\% confidence ranges allowed for the $\Lambda$CDM prediction using all data.}
\label{fig:h0}
\end{center}
\end{figure}

Effective distance $D_V(z)$ is presented in Figure~\ref{fig:DV}. A good agreement is found between data and best fit scenario predictions  for  the $\Lambda$CDM and the  Galileon models. However, as for luminosity distances, the strong variations in the Galileon $w$ parameter do not translate into significant deviations in the $D_V$ distances. The same is observed for angular distance $D_A(z)$, see Figure~\ref{fig:HDA} (left). The reason lies in the fact that these distances are integrals of the Hubble expansion rate $H(z)$.

The impact of the Galileon $w$ variations is more visible directly on the expansion rate (Figure~\ref{fig:HDA} right), which exhibits deviations from $\Lambda$CDM on a larger redshift range than distances. In this plot, the BOSS CMASS anisotropic measurement is in tension with the Galileon best fit scenarios, which predict a lower expansion rate than the $\Lambda$CDM model at $z\gtrsim 0.5$. On the other hand, the $D_A(z=0.57)$ measurement agrees correctly with the Galileon predictions and the impact of that result on the final $\chi^2$ is not too important. 

As for future prospects, Figure~\ref{fig:HDA} gives confidence that distance measurements with sub-percent accuracy will be able to distinguish $\Lambda$CDM from dark energy models with dynamical scalar fields.
The best target range lies between $z\approx0.5$ and $z\approx1.5$, and requires $H(z)$ measurements, i.e. anisotropic BAO observable derivation. 
In conclusion, $H(z)$ measurements in a wide redshift range at several redshifts will be a key probe used to either confirm $\Lambda$CDM or Galileon models. Future surveys such as \textit{Euclid} \citep{bib:euclid}, DESI \citep{bib:desi}, or LSST \citep{bib:lsst} have the necessary precision to do this, as we will show in a future paper.

To distinguish Galileon models from $\Lambda$CDM, $H_0$ direct measurements may provide an additional lever arm. Indeed, the present $H_0$ direct measurements (Table~\ref{tab:h0}), and in particular the last measurement from \cite{bib:riess16}, agree better with the Galileon models than with $\Lambda$CDM (see Tables~\ref{tab:lcdm} -- \ref{tab:resultsc0cG} and Figure~\ref{fig:h0}) even though they were not used in the fitting procedure. If the tension with $\Lambda$CDM is confirmed by future more precise $H_0$ measurements, Galileon models can be favoured.  

\subsubsection{Growth of structure measurements}

\begin{figure}[h]
\begin{center}
\epsfig{figure=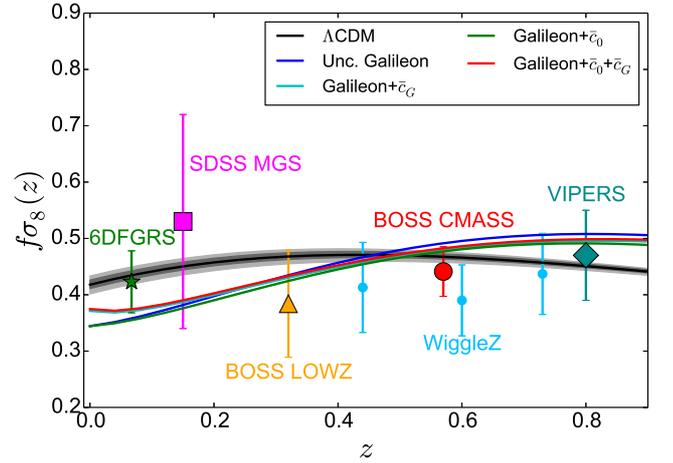, width=\columnwidth} 
\caption[]{Comparison of the best fit Galileon and $\Lambda$CDM $f\sigma_8(z)$ predictions with data. The grey bands show the 68\% and 95\% confidence ranges allowed for the $\Lambda$CDM prediction using all data. The data points are from Table~\ref{tab:gof}.}
\label{fig:fs8}
\end{center}
\end{figure}

Growth of structure measurements are often advocated as the key probe for distinguishing modified gravity theories from the $\Lambda$CDM model. However, they are also the most difficult to use as the growth rate of structures depends on non-linear processes at small scales in general relativity and in modified gravity theories. Thus, the use of growth measurements often relies on linearised equations which are supposed to be valid in the scale range explored in the data (see e.g. \citetalias{bib:neveu} for more details). 

Both \citetalias{bib:neveu} and \citetalias{bib:neveu14} exhibit small tensions between constraints from growth data and distances  in the $\Lambda$CDM and Galileon models. In this paper, the updated growth data set involves measurements less plagued by non-linear effects than those in our previous works. As better agreement is now observed between growth and distance data, it seems that the growth measurements become more and more mature thanks to larger surveys. 

Figure~\ref{fig:fs8} shows the quality of the agreement between growth measurements and the cosmological models. Because of a different gravitational coupling, Galileon models present a different $f\sigma_8(z)$ evolution to that of  $\Lambda$CDM. The largest difference occurs at low redshift, but the volume of galaxy surveys in this range of redshifts is too limited to provide $f\sigma_8$ measurements precise enough to distinguish the Galileon theory from $\Lambda$CDM. However, $f\sigma_8$ measurements at redshifts $z \gtrsim  0.8$ will have this capability. In particular, the precision expected for the DESI, Euclid, and LSST surveys will allow the Galileon growth rate predictions to be tested. However, work remains to be done to check whether the Vainshtein radius, which determines at which scale the Galileon fifth force becomes negligible because of non-linearities, is within the range of scales probed by the observations.

\subsection{External constraints}

The following non-cosmological constraints were not used directly in our analysis to constrain coupled Galileon models. However, we can check a posteriori whether the Galileon best fit scenarios are compatible with these requirements.
 
\subsubsection{Lunar Laser Ranging experiment}

The conformal coupling can be tested using measurements of the Newton constant variations with time. As  shown in \cite{bib:babichev}, the conformal coupling of the Galileon is severely constrained by  solar system tests. Indeed, from equation~\ref{eq:00}, the Newton constant in a conformally coupled Galileon model can be redefined as
\begin{equation}
G_N  = G_N^0  / (1-2\bar c_0 \bar y) ,
\end{equation}
where $G_N^0$ is the present Newton constant. Measurements from the Lunar Laser Ranging (LLR) experiments \citep{bib:llr} put tight constraints on the variation of $G_N$
\begin{equation}
\frac{\dot{G}_N}{G_N} < 1.3 \times 10^{-12}\, \mathrm{yr}^{-1} \approx 0.02 H_0 \Leftrightarrow \frac{G_N'}{G_N} < 0.02.
\end{equation}
In the Galileon theory, we have $G_N' / G_N (z=0) = -2 \bar c_0 \y'_0 = -2 \bar c_0$ to be compared with $G_N' / G_N (z=0) < 0.02$. Using the conformally coupled best fit Galileon model from Table~\ref{tab:resultsc0} (last row), we have $\bar c_0 = -0.013 \pm 0.008$ and thus $G_N'/G_N(z=0) = 0.026 \pm 0.016$, which is compatible with the LLR constraint. 

\subsubsection{Propagation of gravitational waves on cosmological scales}

\begin{figure}[hbtp]
\begin{center}
\epsfig{figure=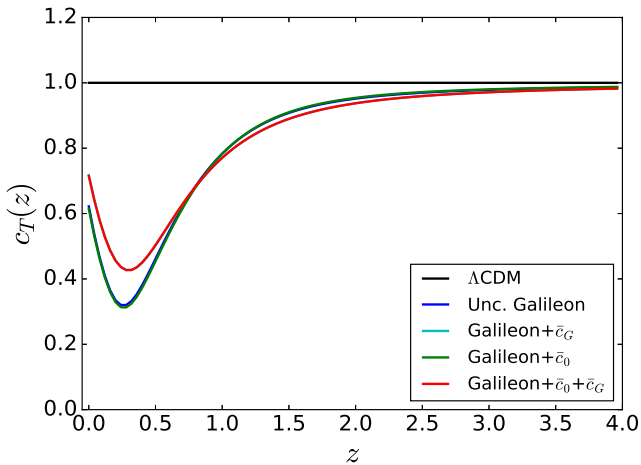, width=\columnwidth} 
\caption[]{Gravitational wave speed of propagation $c_T(z)$ as a function of redshift in the best fit scenarios of the four Galileon models studied in this paper, compared with the general relativity prediction $c_T=1$.} 
\label{fig:cT}
\end{center}
\end{figure}

In modified gravity models the gravitational wave speed of propagation $c_T$ can be different from the speed of light. For the Galileon theory studied in this paper, $c_T$ is defined in equation~\ref{eq:cs2}. For the four Galileon models tested in this paper, the best fitting scenarios give $c_T(z=0) \approx 0.7$ and the propagation is subluminal all over the past history of the Universe\footnote{If superluminal, the graviton may emit Cherenkov light that can be observed by telescopes \citep{bib:brax16}.}. Figure~\ref{fig:cT} shows the redshift evolution of $c_T$ in the four best fit scenarios obtained in this paper.

Attempts to constrain the gravitational wave speed have been conducted for many years, but most of them stand on astrophysical scales where the Galileon gravity is in the Vainshtein regime, whereas our prediction of $c_T$ from~\ref{eq:cs2} is valid at cosmological scales. Constraints on $c_T$ have been derived using the Hulse-Taylor binary pulsar assuming a Vainshtein screening, but this still relies on many assumptions (see \citealt{bib:jimenez}). 
The propagation speed of gravitational waves can also be constrained with the gravi-Cherenkov effect \citep{bib:moore}. This effect predicts that the ultra-relativistic cosmic rays should emit gravitons if they travel faster than the speed of gravity. The predicted loss of energy through this emission, together with the observation of cosmic rays at the highest energy, bring constraints on $c_T$, depending on the source distance. If the source is galactic, then the Vainshtein screening mechanism is at play and the constraint \citep{bib:moore} cannot be directly applied to the cosmological predictions we performed. Moreover, as no extragalactic source of ultra-relativistic cosmic rays has been identified yet, the most energetic cosmic rays that have been detected are likely to come from a source located within our local galaxy group but not at hundreds of megaparsecs.

Bounds coming from observations of gravitational wave sources at cosmological scales would be safer when working with cosmological models. The impact of modifications of $c_T$ has been searched for in the B-mode polarisation spectrum using BICEP2 data, but no firm conclusions could be derived owing to insufficient statistics \citep{bib:amendola,bib:raveri}. Data coming from future B-mode polarisation spectrum measurements may bring tight constraints \citep{bib:amendola,bib:raveri}.

The recently discovered gravitational wave signal GW150914 \citep{bib:ligo2016} could tightly constrain the Galileon theory on cosmological scales through the  measurement of the graviton speed, if a firm electromagnetic counterpart were found. For instance, with $c_T\approx 0.7$, if some light has been produced by the black hole merger, the electromagnetic signal is expected to arrive well before the gravitational wave. In the case of GW150914, detected at $1.3\, \mathrm{Gly}$, light emitted at the moment of the black hole fusion would have arrived $\approx 5\times10^8\, \mathrm{yr}$ before the gravitational signal.

 Even if a light or neutrino counterpart is not expected a priori for such a black hole merger, a broadband follow-up has been conducted by many observatories \citep{bib:ligo-followup}. At the moment no firm electromagnetic counterpart to GW150914 has been  found to set a direct measurement of $c_T$. Although the \textit{Fermi} Gamma-ray space telescope has reported a detection compatible in time and space with the GW150914 event, this weak transient can also be a coincidence \citep{bib:fermi}. In the next years, future detections of extragalactic black hole mergers or even neutron star mergers (for which we do expect a light emission) with ground-based gravitational wave interferometers will bring strong conclusions on the behaviour of gravity on cosmological scales. 

\subsubsection{Particle physics and cosmology}

Particle physics may be used to set constraints on the disformal coupling between matter and scalar fields \citep{bib:brax15}. In particular, the Compact Muon Solenoid (CMS) collaboration has recently published experimental limits on the Branon theory \citep{bib:monophoton}. This extra-dimensional model exhibits massive scalar fields scaling the 4D brane fluctuations into a broader space, which depends of the brane energy scale tension $f$ \citep{bib:dobado,bib:cembranos2004}. The scalar particles $\pi_B$ are possible dark matter candidates and are coupled to Standard Model (SM) fields in the form $(\partial_\mu \pi_B \partial_\nu \pi_B - M^2_B \pi_B \pi_B g_{\mu\nu}/4)T^{\mu\nu}_{SM}/2f^4$, with $M_B$ the Branon mass \citep{bib:cembranos2004}. In the case of a massless Branon, the Branon coupling to matter can be identified with the Galileon disformal coupling in the Einstein frame. Providing that the transformations between the Einstein and Jordan frames and between the classical and quantum levels 
can be established thoroughly and the non-trivial background value of $\pi$ is taken into account accordingly, the CMS 95\% confidence level (CL) experimental limit $f > 412\,$GeV \citep{bib:monophoton} could be translated into an upper bound on the $c_G$ coupling. 

\subsection{Tracker solution of the uncoupled Galileon model and comparison with \cite{bib:barreira14}}

As shown in the previous sections, Galileon models can provide a good alternative to $\Lambda$CDM to describe present cosmological data. Combining cosmological fits and external constraints, the uncoupled Galileon model appears to be the most promising scenario to consider. 

In a recent phenomenological work on the uncoupled Galileon model  \cite{bib:barreira14} has suggested  considering only a tracker solution as only scenarios that asymptotically reach the tracker solution well before the onset of the accelerated expansion era can provide a reasonable fit to CMB data. Then only this subspace of the full parameter space of the uncoupled Galileon model would be interesting to explore. The same paper also shows that the tracker solution is rejected by present cosmological data using the full CMB power spectrum and BAO measurements but no supernova or growth data. A good agreement with data can be retrieved if massive neutrinos are added to the fit \citep{bib:barreira14b}.

Our findings that the general uncoupled Galileon model provides good agreement with data seems in contradiction with the above results. It may invalidate the assumption that the tracker solution is representative of the only solutions that provide a good fit to CMB data. The different result could also stem from differences in methodology or data sets. To check this point, we also tested the tracker solution of the uncoupled Galileon model within our framework, as described in the next sections.

\subsubsection{Uncoupled Galileon tracker solution}

\begin{table*}[htb]
\caption[]{Best fit values from different data samples for the tracker solution of the uncoupled Galileon model.}
\label{tab:results_tracker}
\begin{center}
\begin{tabular}{cccccccccc} \hline \hline \\ [-1ex]
Probe & $\Omega_m^0$ & $\bar c_2$ & $\bar c_3$ & $h$ & $\Omega_b^0h^2$ & $ \chi^2$  & $ \chi^2_{\Lambda CDM}$  & $ \chi^2_{Unc.}$  & $N_{\rm data}$ \\  [1ex] \hline \\ [-1ex]
SNe~Ia & $0.392^{+0.038}_{-0.034}$ & $-3.9^{+1.9}_{-2.8}$ & $-0.8^{+1.3}_{-1.6}$ & - & - & 693.5 & 691.0 & 692.8 & 740 \\ [1ex]  \hline \\ [-1ex] 
Growth & $0.213^{+0.057}_{-0.041}$ & $-5.6^{+1.2}_{-2.5}$ & $-1.6^{+0.6}_{-1.4}$ & - & - & 2.8 & 2.9 & 2.9 & 12 \\ [1ex]  \hline \\ [-1ex] 
\textit{Planck}+BAO+Ly$\alpha$ & $0.261^{+0.006}_{-0.005}$ & $-5.3^{+2.1}_{-3.1}$ & $-1.1^{+1.8}_{-1.7}$ & 0.763 & 0.0237 & 43.3 & 14.5 & 22.0 & 15 \\ [1ex]  \hline  \\ [-1ex]  
All & $0.264^{+0.006}_{-0.005}$ & $-5.0^{+1.1}_{-1.7}$ & $-1.7^{+0.4}_{-1.1}$ & 0.760 & 0.0237 & 754.6 & 710.6 & 731.9 & 767 \\ [1ex]  \hline  \\ [-1ex]  
\end{tabular}
\tablefoot{The JLA SN~Ia sample is used with systematics included; $\alpha$ and $\beta$ are fixed to their marginalised values. $h$ and $\Omega_b^0h^2$ have been minimised so no uncertainties are provided. Best fit $\chi^2$ values from $\Lambda$CDM and uncoupled Galileon models are reported for comparison.}
\end{center}
\end{table*}

\begin{figure*}[hbtp]
\begin{center}
\epsfig{figure=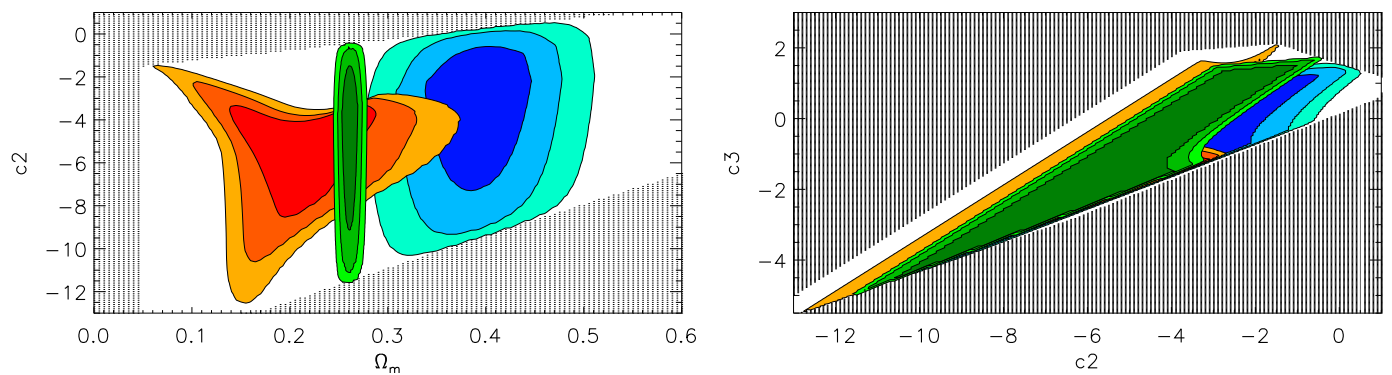, width=\textwidth} 
\caption[]{Cosmological constraints on the tracker solution of the uncoupled Galileon model from growth data (red), JLA (blue), and \textit{Planck}+BAO+Ly$\alpha$ data (green). The filled dark, medium, and light yellow contours enclose 68.3, 95.4, and 99.7\% of the probability, respectively. Dark dotted regions correspond to scenarios rejected by theoretical constraints.}\label{fig:gof_all_combined_tracker}
\end{center}
\end{figure*}

The existence of an attractor solution to the uncoupled Galileon equations was originally proved in \cite{bib:felice2010} and  \cite{bib:felice2011}. The authors showed that, whatever the initial conditions for the Galileon field at the Big Bang, the Galileon equations converge towards a solution where $H \dot{\pi}$ is a constant. The tracker solution has been widely tested with cosmological data in previous studies (see e.g. \cite{bib:nesseris}, \cite{bib:barreira14} and \cite{bib:barreira14b}).

In the parametrisation frame we used, the tracker solution is characterised by
\begin{equation}\label{eq:tracker}
\h^2 \x = 1.
\end{equation}
Multiplying equation~\ref{eq:00} by $\h^2$ and inserting equation~\ref{eq:tracker}, we obtain
\begin{equation}\label{eq:OOtracker}
\h^4 =\left(\frac{\Omega_m^0}{a^3}+\frac{\Omega_r^0}{a^4}\right)\h^2 + \frac{\bar  c_2}{6}  - 2\bar  c_3  + \frac{15}{2}\bar c_4 - 7 \bar c_5
\end{equation}
in the uncoupled case. With $\Omega_\pi^0 = \frac{\bar  c_2}{6}  - 2\bar  c_3  + \frac{15}{2}\bar c_4 - 7 \bar c_5$, this leads to an analytical formula to compute the expansion of the Universe corresponding to the tracker solution:
\begin{align}
\h^2 & = \frac{1}{2}\left[\frac{\Omega_m^0}{a^3}+\frac{\Omega_r^0}{a^4} + \sqrt{4\Omega_\pi^0+\left(\frac{\Omega_m^0}{a^3}+\frac{\Omega_r^0}{a^4}\right)^2}\right]\label{eq:htracker}, \\ 
\x & = 1 / \h^2.\label{eq:xtracker}
\end{align}

Equation~\ref{eq:tracker} can also translate into a new constraint on the $\bar c_i$ parameters. Indeed, at $z=0$ the Galileon field equation of motion is
\begin{equation}
\bar c_2 - 6 \bar c_3 +18 \bar c_4 - 15 \bar c_5 = 0,
\end{equation}
for the tracker solution. So, combining this equation with the constraint equation~\ref{eq:c5} (which comes from the (00) Einstein equation at $z=0$), we can fix two $\bar c_i$ parameters. We choose to fix $\bar c_4$ and $\bar c_5$:
\begin{align}
\bar c_4 &  = \frac{1}{9}\left[10\left(\Omega_m^0 + \Omega_r^0 -1 \right) - 3 \bar c_2 + 8 \bar c_3 \right], \\
\bar c_5 &  = \frac{1}{3}\left[4\left(\Omega_m^0 + \Omega_r^0 -1 \right) - \bar c_2 + 2 \bar c_3 \right].
\end{align}
The set of equations presented in this section are equivalent to the tracker equations exposed in \cite{bib:barreira14} and \cite{bib:brax15}. In summary, using the Galileon tracker solution has remarkable advantages on the computational side, since one more parameter can be fixed and analytical solutions are provided. In the following, to compute the evolution of Galileon scenarios in the tracker solution, we took advantage of these analytical formula instead of resorting to the numerical integration of equations~\ref{eq:dpi} and \ref{eq:dh}.

\subsubsection{Cosmological constraints}

Cosmological constraints on the uncoupled Galileon tracker model are presented in Table~\ref{tab:results_tracker} and Figure~\ref{fig:gof_all_combined_tracker}, using the same data as in previous sections.

Figure~\ref{fig:gof_all_combined_tracker} reveals that the constraints from supernovae and \textit{Planck}+BAO+Ly$\alpha$ data disagree at 3$\sigma$ and that the growth data contour barely overlaps that from supernovae.
When compared to values obtained in the $\Lambda$CDM or general uncoupled Galileon model, the $\chi^2$ value of the tracker solution is significantly worse for \textit{Planck}+BAO+Ly$\alpha$ data, resulting in a degraded global  $\chi^2$ value (see Table~\ref{tab:results_tracker}).

In conclusion, the tracker solution of the uncoupled Galileon model cannot provide a good fit to all cosmological data and thus is excluded. On this point we agree with \cite{bib:barreira14}. The hypothesis that the Galileon follows the tracker solution all along the history of the Universe is excluded by cosmological data. But this does not mean that the Galileon theory is in trouble. As shown in section~\ref{sec:results}, Galileon models not restricted to tracker solutions have been proved to be compatible with present cosmological data, using CMB priors only, even if the scenarios converge to tracker solutions at late times. It remains to be checked whether they are also compatible with the full CMB power spectrum, which will be the subject of a future work (see Appendix~\ref{sec:convergence} for further discussions).

\section{Conclusion}\label{sec:concl}

We have compared the $\Lambda$CDM and Galileon models to the most recent cosmological data from SNe~Ia, CMB, BAO, and growth rate measurements. The uncoupled Galileon case and  the Galileon models disformally and conformally coupled to matter were tested. Compared to our previous publication (\citetalias{bib:neveu14}), the data set was updated using the most recent results for CMB (in the form of distance priors), BAO, and growth data.

All probes agree well when compared to the $\Lambda$CDM model, and the fit is mainly driven by the CMB priors and the BAO measurements. In the Galileon case, the CMB priors set precisely the $\Omega_m^0$ best fit value with unprecedented precision, but the constraints on the $\bar c_i$ parameters are driven by the combination of all probes. A good agreement with data is observed with each of the probes for all models. 

We provided the first cosmological constraints on the conformal coupling parameter in the framework of the Galileon model. We showed that this type of coupling is not favoured by cosmological data and that the disformally coupled model is preferred against the other Galileon cases, with a non-zero coupling excluded at the $2.3\sigma$ level. We also showed  how non-cosmological data sets can bring constraints on the Galileon parameters. 

\cite{bib:barreira14} used the full CMB power spectrum to set constraints on the tracker solution of the uncoupled Galileon theory and found some tensions between CMB and BAO constraints. We also tested  the uncoupled Galileon tracker solution and found incompatibilities with supernovae and CMB+BAO data. 
Although its tracker solution was rejected by the data, the uncoupled Galileon model not restricted a priori to any particular type of solution provides as good an agreement with cosmological data as the $\Lambda$CDM model. 

The Galileon theory provides specific predictions on the expansion and growth histories of the Universe that can be probed by future dark energy experiments. But non-linearities of the Vainshtein mechanism may prevent growth predictions to be compared with future precise growth measurements from DESI, LSST, or Euclid. However, we argued that upcoming distance measurements can be important in order to discriminate between the Galileon and the $\Lambda$CDM models. This will be detailed in a forthcoming paper.

\begin{acknowledgements} 

We thank Philippe Brax for fruitful discussions and suggestions on the conformally coupled Galileon model. Part of this research was conducted by the Australian Research Council Centre of Excellence for All-sky Astrophysics (CAASTRO), through project number CE110001020.

\end{acknowledgements}

\appendix

\section{Computation of $r_d$}\label{sec:rd}

Since the publication of the first \Planck cosmological results \citep{bib:planck}, the \cite{bib:eisenstein98} fitting formula to compute $z_d$ is no longer precise enough to describe data and should be avoided. The use of the CAMB \citet{bib:camb} non-approximate computation of $r_d$ is now necessary. However, it has been shown in \cite{bib:mehta} that the ratio $r_d^{\rm fid}/r_d$ is still  independent of the methodology used to compute $r_d$. As emphasised by \cite{bib:boss14},    any of these conventions for the sound horizon can be followed when using their measurements as long as   consistency is maintained when evaluating $r_d$ and $r_d^{\rm fid}$.

We checked that our code provided $r_d^{\rm fid}/r_d$ ratios identical to those of CAMB. We explored both flat and open $\Lambda$CDM model parameter spaces and computed ratios with CAMB and our code. In the latter, $z_d$ is computed using the \cite{bib:eisenstein98} formula and $r_d$ by numerically computing the integral
\begin{equation}
r_d=r_s(z_d)\frac{H_0}{c}=\int_0^\frac{1}{1+z_d}da\frac{\bar c_s(a)}{a^2\bar H(a)} 
\end{equation}
and $\bar c_s$ is the usual normalised sound speed in the baryon-photon fluid before recombination
\begin{equation}
\bar c_s = \frac{1}{\sqrt{1+3(3\Omega_b^0/4\Omega_\gamma^0)a}},
\end{equation}
where $\Omega_b^0$ is the current baryon energy density parameter. 

The fiducial cosmology chosen to make this test is $\Omega_m^0=0.27$, $\Omega_\Lambda^0=0.73$, $h=0.7$, and $\Omega_b^0h^2=0.0224$. We found $r_{d,\rm CAMB}^{\rm fid} = 149.74$\,Mpc and $r_{d,\rm Cosfitter}^{\rm fid} = 153.63$\,Mpc. As pointed out in \cite{bib:boss14} and \cite{bib:planck}, the discrepancy is  about 2.5\% and CAMB is better able  to reproduce \Planck data. However, when exploring a wide part of the $\Lambda$CDM parameter space ($0 < \Omega_m^0 < 0.6$ and $0 < \Omega_\Lambda^0 < 1.4$, with $h$ and $\Omega_b^0h^2$ varied thanks to the minimisation procedure), the $r_{d,\rm Cosfitter}^{\rm fid}/r_{d,\rm Cosfitter}$ ratios differ from those of CAMB by at most $0.6\%$ (and by only $0.12\%$ around the best fit values). Compared to the $\approx 2\%$ uncertainties from the BAO measurements, this potential systematic uncertainty is thus negligible and our code is precise enough to evaluate $r_{d}^{\rm fid}/r_{d}$.

We also tested the new approximate formula for $r_d$ from \cite{bib:aubourg} equation 16. This formula reproduces CAMB $r_d$ ratios nearly exactly around the $\Lambda$CDM best fit values, but can differ by $7\%$ in some points of the same parameter space previously explored.

\section{Goodness of fit}\label{sec:goodness}

\begin{table*}[h!]
\begin{center}
\caption{$p$-values of the compatibility test between a normal distribution of mean $0$ and variance $1$ and the residual distribution 
obtained for the $\Lambda$CDM model, and the general and tracker solutions of the uncoupled Galileon model. The best fit $\chi^2$ values are shown between parentheses.}\label{tab:pvalues}
\begin{tabular}{ccccc}\hline\hline\\ [-1ex] 
Probes & $N_{\rm data}$ &  $\Lambda$CDM & Unc. Galileon & Tracker\\ [1ex] \hline \\[-1ex]
                         SNe Ia &            740 &        0.38 (691.0) &         0.21  (692.8)&   0.23 (693.5)\\
                         Growth &             12 &         0.17 (2.9) &         0.17 (2.9)&   0.26 (2.8) \\
 \textit{Planck}+BAO+Ly$\alpha$ &             15 &        0.91 (14.5) &          0.85 (22.0) &  0.08 (43.3)\\
                            All &            767 &        0.23 (710.6) &         0.71 (731.9) &   0.78 (754.6)\\
[1ex]  \hline
 \end{tabular}
\end{center}
\end{table*}

To check the goodness of a cosmological fit when the number of degrees of freedom $N_{\rm dof}$ is not available, an easy way is to compare the number of measurements $N_{\rm data}$ with the obtained $\chi^2$. When both are equivalent, it means that residuals $(y^{\rm i}_{\rm mes} - y^{\rm i}_{\rm mod})/\sigma^{\rm i}_y$, where $y^{\rm i}_{\rm mod}$ are predictions for observable $y$ compared to measurements $y^{\rm i}_{\rm mes}$ with uncertainties $\sigma^{\rm i}_y$, follow a Gaussian distribution of mean $0$ and variance $1$. 

To have a more quantitative statement, it is possible to  test whether the observed residuals are likely to come from a normal distribution of mean $0$ and variance $1$ by using a Kolmogorov-Smirnov test, as suggested in \cite{bib:andrae}. The output of this test is a $p$-value which gives the probability that the observed data are drawn from the probed model. 
We performed the exercise with our best fit scenarios for the $\Lambda$CDM, uncoupled Galileon, and tracker Galileon models. The $p$-values are reported in Table~\ref{tab:pvalues} separately for each probe and their combination. We note  that we did not take into account the covariances between the measurements for this test (we took only the diagonal terms).

Table~\ref{tab:pvalues} shows that the higher the $p$-value, the closer the $\chi^2$ is to $N_{\rm data}$. A low $p$-value is obtained in all other cases, whether the $\chi^2$ is greater than $N_{\rm data}$, which may reveal tensions between data and predictions, or lower than $N_{\rm data}$, which may indicate an excellent fit to data or point towards overestimated uncertainties.

The conclusions based on $p$-values are similar to those quoted in the paper when comparing the best fit $\chi^2$ with $N_{\rm data}$. 
The tracker Galileon agrees with all data combined ($p$-value of 0.78), but this masks a rejection by \textit{Planck}+BAO+Ly$\alpha$ data ($p$-value of 0.08). The $\Lambda$CDM provides a good fit to all data separately and an excellent fit to all data combined ($p$-value of 0.23). The general uncoupled Galileon model offers a good fit to all probes separately and combined ($p$-value of 0.71) and thus remains a robust alternative to $\Lambda$CDM. The above test confirms that comparing $\chi^2$ values with $N_{\rm data}$ provides a reasonable estimate of the goodness of fit for models studied in that work.

\begin{figure*}[hbtp]
\begin{center}
\epsfig{figure=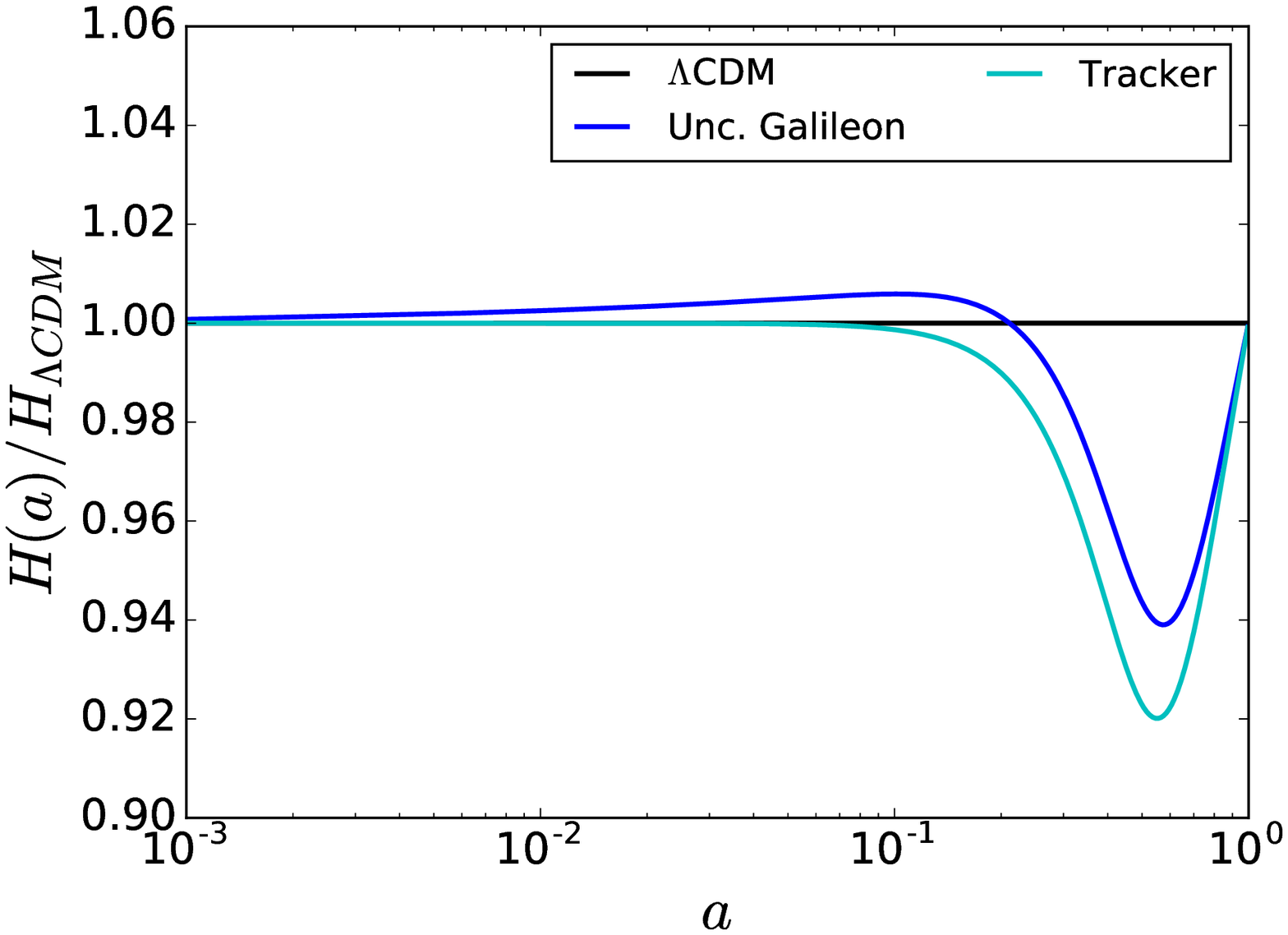, width=0.45\textwidth} 
\epsfig{figure=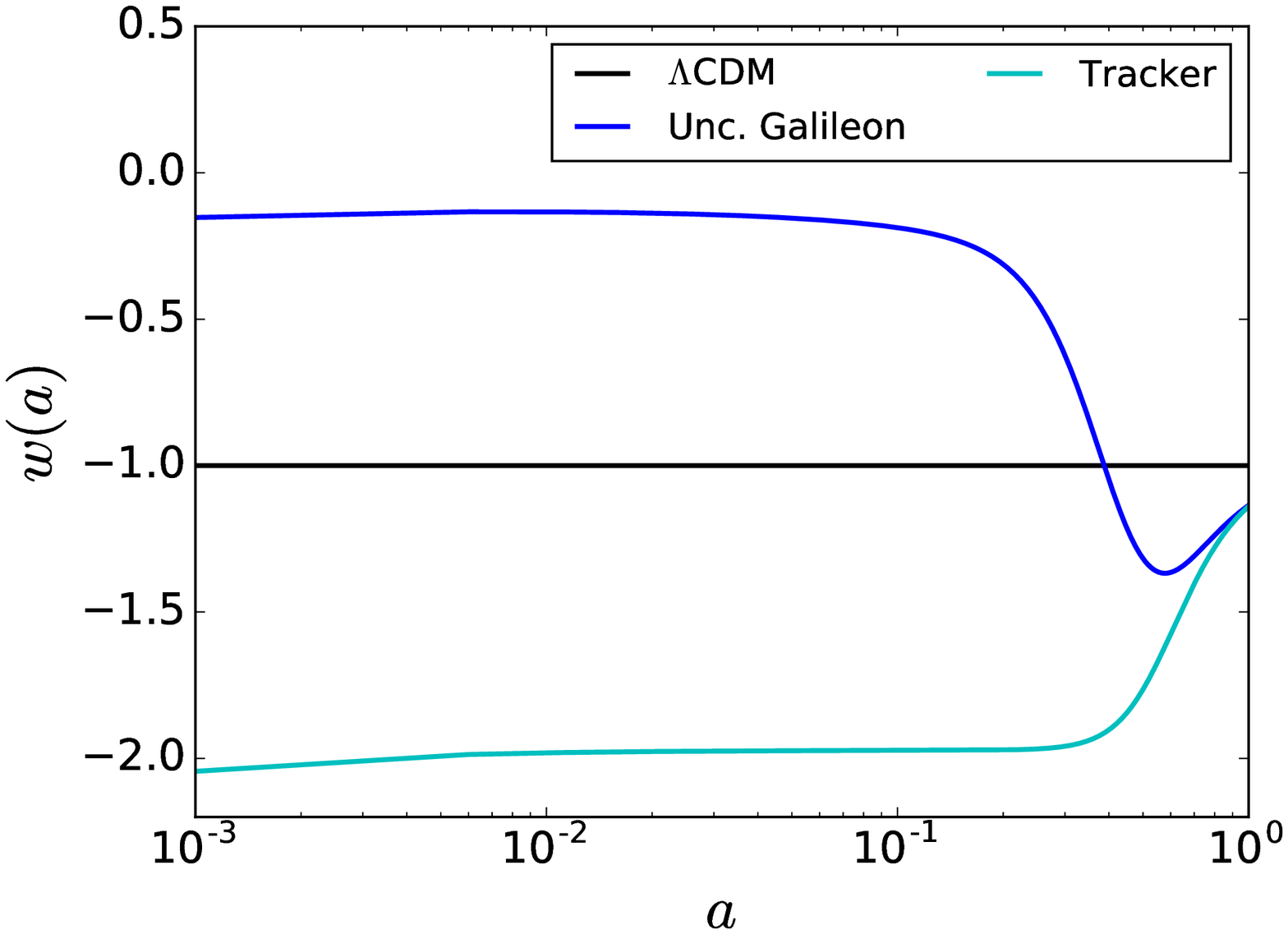, width=0.45\textwidth} 
\caption[]{Evolution of the Hubble parameter $H(a)$ (left) and the state equation parameter $w(a)$ (right) for the uncoupled Galileon best fit scenario, the asymptotic tracker solution, and the $\Lambda$CDM model with $\Omega_m^0$ and $\Omega_r^0$ set to the Galileon best fit values and $\Omega_\Lambda^0=1-\Omega_m^0-\Omega_r^0$.}\label{fig:tracker}
\end{center}
\end{figure*}

\section{Convergence to the tracker solution}\label{sec:convergence}

In this paper we showed that the uncoupled Galileon model is compatible with cosmological data, using CMB priors but not the full CMB power spectrum. \cite{bib:barreira14c},  Appendix A, argued that only Galileon scenarios that reach their asymptotic tracker solution ``sufficiently early'' provide a reasonable fit to the low-$\ell$ part of the power spectrum. In order to check whether this is the case for our best fit scenario, we provide in Figure~\ref{fig:tracker} the $\h(z)$ and $w(z)$ evolutions for the uncoupled Galileon best fit scenario and its associated tracker solution. The latter is computed assuming $\Omega_{\pi}^0 = 1 - \Omega_m^0 - \Omega_r^0$ with $\Omega_m^0$ and $\Omega_r^0$ the best fit values of the associated uncoupled Galileon scenario, from equations~\ref{eq:htracker} and \ref{eq:xtracker}.

In Figure~\ref{fig:tracker} we plot the evolution of $H(a)$ for the uncoupled Galileon best fit scenario, the associated tracker solution, and a $\Lambda$CDM scenario. The latter is set to have the same best fit values as the uncoupled Galileon best fit scenario for the $\Omega_m^0$ and $\Omega_r^0$ parameters, and $\Omega_\Lambda^0=1-\Omega_m^0-\Omega_r^0$. In this way, all the scenarios have the same amount of matter, radiation, and dark energy at $a=1$, but the nature of dark energy is different. As we can see both Galileon scenarios depart from the $\Lambda$CDM model in the late universe, but at the level of a few percent only. In the uncoupled Galileon scenario, $H(a)$ is greater than the Hubble parameter of the associated $\Lambda$CDM model from decoupling to $a\approx 0.2$ and then goes to negative values. Concerning the tracker solution, $H(a)$ is always lower than in the $\Lambda$CDM model and explores lower expansion rate values than in the Galileon best fit scenario. This difference explains that observational constraints using cosmological distances rule out the tracker solution, but not the full Galileon model.
 
Figure~\ref{fig:tracker} shows the evolution of $w$ for the uncoupled Galileon best fit scenario, the associated tracker solution, and the $\Lambda$CDM model. If we compare this result with what is published in \cite{bib:barreira14c}, it seems that the uncoupled Galileon best fit scenario reaches the tracker solution too late ($a\gtrsim 0.6$). However, \cite{bib:barreira14c} studied the cubic Galileon model ($\bar c_4 = \bar c_5=0$), and  their best fit $w$ values go from $+0.2$ to $-2$. In Figure~\ref{fig:tracker} the $w$ values of the best fit Galileon scenario remain between $-0.2$ and $-1.3$, which is likely to correspond to different physics. This indicates that the full Galileon model may be incompatible with the low-$\ell$ part of the CMB power spectrum according to the \cite{bib:barreira14c} arguments. But, as the two studies have different methodologies in particular concerning the Galileon initial conditions, applying these arguments to our study may not be direct. In a future work we plan to have CMB power spectrum predictions for the full uncoupled Galileon model and to compare them with the most recent \textit{Planck} data with our methodology.

\end{document}